\documentclass[12pt]{iopart}
\pdfoutput=1

\usepackage{graphicx}
\usepackage{subfigure}
\usepackage[colorlinks=true,urlcolor=blue,citecolor=blue]{hyperref}
\usepackage{doi}

\expandafter\let\csname equation*\endcsname\relax
\expandafter\let\csname endequation*\endcsname\relax

\usepackage{amsfonts}
\usepackage{amsmath}
\usepackage{amssymb}
\usepackage{amstext}
\usepackage[numbers,sort&compress]{natbib}

\begin{document}

\title[Fractional Random walks with absorbing walls]{Probability density of fractional Brownian motion
and the fractional Langevin equation with absorbing walls}

\author{Thomas Vojta and Alex Warhover}
\address{Department of Physics, Missouri University of Science and Technology, Rolla, MO 65409, USA }

\vspace{10pt}
\begin{indented}
\item[] \today
\end{indented}

\begin{abstract}
Fractional Brownian motion and the fractional Langevin equation are models of anomalous diffusion processes
characterized by long-range power-law correlations in time. We employ large-scale computer simulations
to study these models in two geometries, (i) the spreading of particles on a semi-infinite domain with an
absorbing wall at one end and (ii) the stationary state on a finite interval with absorbing boundaries at
both ends and a source in the center. We demonstrate that the probability density and other properties of the
fractional Langevin equation can be mapped onto the corresponding quantities of fractional Brownian motion
driven by the same noise if the anomalous diffusion exponent $\alpha$ is replaced by $2-\alpha$.
In contrast, the properties of fractional Brownian motion and the fractional Langevin equation with \emph{reflecting}
boundaries were recently shown to differ from each other qualitatively. Specifically, we find that the
probability density close to an absorbing wall behaves as $P(x) \sim x^\kappa$ with the distance $x$ from
the wall in the long-time limit. In the case of fractional Brownian motion, $\kappa$ varies with
the anomalous diffusion exponent $\alpha$ as $\kappa=2/\alpha -1$, as was conjectured previously. We also
compare our simulation results to a perturbative analytical approach to fractional Brownian motion.
\end{abstract}


%\submitto{\JSTAT}

\tableofcontents
\markboth{}{}

%%%%%%%%%%%%%%%%%%%%%%%%%%%%%%%%%%%%%%%%%%%%%%%%%%%%%%%%%%%%%%%%%%%%%%%
\section{Introduction}
\label{sec:introduction}
%%%%%%%%%%%%%%%%%%%%%%%%%%%%%%%%%%%%%%%%%%%%%%%%%%%%%%%%%%%%%%%%%%%%%%%

Within the stochastic approach spearheaded by Einstein \cite{Einstein_book56}, Smoluchowski
\cite{Smoluchowski18}, and Langevin \cite{Langevin08}, diffusion is commonly understood as
random motion. If this motion is local in time and space in the sense that (i) it features
a finite correlation time after which individual steps become statistically independent, and
(ii) the displacements over a correlation time feature a finite second moment, the central limit
theorem applies to the sequence of random steps.
This leads to the well-known linear relation $\langle x^2 \rangle \sim t$
between the mean-square displacement of the moving particle and the elapsed time $t$
that characterizes normal diffusion \cite{Hughes95}.

Anomalous diffusion, i.e., random motion whose mean-square displacement does not obey the linear
$\langle x^2 \rangle \sim t$ relation, has attracted considerable attention in
recent years (for reviews see, e.g., Refs.\
\cite{MetzlerKlafter00,HoeflingFranosch13,BressloffNewby13,MJCB14,MerozSokolov15,MetzlerJeonCherstvy16}
and references therein), partially because novel microscopic techniques provide access to the trajectories
of single molecules in complex environments \cite{XCLLL08,BrauchleLambMichaelis12,ManzoGarciaParajo15}.
Anomalous diffusion can be caused by various mechanisms that violate the condition of
locality in space and time. For example, long-range correlations in time between individual
random displacements (steps) can produce subdiffusion (for which $\langle x^2 \rangle$ grows
slower than $t$) or superdiffusion (for which $\langle x^2 \rangle$ grows faster than $t$)
even if the step lengths and waiting times are narrowly distributed.
Fractional Brownian motion (FBM) and the fractional Langevin equation (FLE)
are two paradigmatic mathematical models for stochastic processes with long-time correlations.

FBM is a non-Markovian self-similar Gaussian stochastic process with stationary power-law correlated
increments. Its mean-square displacement fulfills the relation $\langle x^2 \rangle \sim t^\alpha$
where $\alpha$ is the anomalous diffusion exponent.\footnote{In the mathematical literature, the Hurst exponent $H=\alpha/2$
is often used instead of $\alpha$.}
If the increments are positively correlated (persistent), the resulting motion is superdiffusive ($\alpha>1$)
whereas anticorrelated (antipersistent) increments produce subdiffusive motion ($\alpha < 1$).
FBM has been employed to model a variety of system ranging from diffusion inside
biological cells \cite{SzymanskiWeiss09,MWBK09,WeberSpakowitzTheriot10,Jeonetal11,JMJM12,Tabeietal13}, the dynamics of
polymers \cite{ChakravartiSebastian97,Panja10}, electronic network traffic \cite{MRRS02}, and the geometry of
serotonergic fibers in vertebrate brains \cite{JanusonisDeteringMetzlerVojta20}, to
fluctuations of financial markets \cite{ComteRenault98,RostekSchoebel13}.
FBM was put forward by Kolmogorov \cite{Kolmogorov40} as well as Mandelbrot and van Ness \cite{MandelbrotVanNess68}.
It has received attention in the mathematical literature (see, e.g., Refs.\ \cite{Kahane85,Yaglom87,Beran94,BHOZ08})
but only limited results are available for FBM in confined geometries because the method of images
\cite{Redner_book01}, typically used for Brownian motion, fails and a generalized diffusion equation
for FBM has not yet been found. Available results include the solution of the
first-passage problem on a semi-infinite interval \cite{HansenEngoyMaloy94,DingYang95,KKMCBS97,Molchan99}),
a conjecture for a two-dimensional wedge domain \cite{JeonChechkinMetzler11}, and results for parabolic
domains \cite{AurzadaLifshits19}.

FBM can be understood as random motion governed by \emph{external} noise \cite{Klimontovich_book95}.
It does not obey the fluctuation-dissipation theorem \cite{Kubo66} and generally does not reach a thermal
equilibrium state. To describe anomalous diffusion in thermal equilibrium, one can use the fractional Langevin
equation \cite{Lutz01}, a generalization of the well-know Langevin equation \cite{Langevin08} involving a
long-range correlated random force and a non-local damping force that fulfill the fluctuation-dissipation
theorem \cite{Kubo66}.

Recent computer simulations of FBM with reflecting walls have shown that the interplay between the long-time
correlations and the geometric confinement affects the probability density function $P(x,t)$ of
the diffusing particle, and leads to strong accumulation and depletion effects.
If the motion is restricted to a one-dimensional semi-infinite interval by a
reflecting wall at the origin, the probability density develops a power-law singularity,
$P \sim x^\kappa$ at the wall \cite{WadaVojta18,WadaWarhoverVojta19}.
The exponent $\kappa$ was conjectured to depend on the anomalous diffusion exponent
via $\kappa=2/\alpha -2$. Particles accumulate at the wall, $\kappa<0$ for persistent noise but are depleted close
to the wall, $\kappa > 0$ for anti-persistent noise. Similarly, simulations of FBM on a finite interval with
reflecting walls at both ends, have shown that the stationary probability density reached for long times is
nonuniform \cite{Guggenbergeretal19,VHSJGM20},
whereas the corresponding distribution for normal diffusion would be uniform. Simulations in
higher dimensions \cite{VHSJGM20} demonstrate analogous behavior.
In contrast, the FLE with reflecting walls features weaker
accumulation and depletion effects \cite{VojtaSkinnerMetzler19}.
The stationary distribution of the FLE on a finite interval is, in fact, completely
uniform independent of the value of $\alpha$, as required in thermal equilibrium.
Some accumulation or depletion of particles (compared to normal diffusion) does occur in nonequilibrium situations such as
the motion on a semi-infinite interval. However, it is less pronounced than in the FBM case.

It is therefore interesting to compare the probability densities of FBM and the FLE in the presence of
absorbing (rather than reflecting) boundary conditions.
The purpose of the present paper is to explore this question systematically by means of large-scale computer simulations.
For the case of FBM, important results have already been obtained in the literature \cite{ChatelainKantorKardar08}.
Specifically, the probability density close to an absorbing wall has been conjectured to vanish as
$P(x) \sim x^\kappa$ with exponent $\kappa =2/\alpha-1$ with the distance $x$ from the wall \cite{ZoiaRossoMajumdar09}.
A perturbative expansion about the normal diffusion case agrees with this conjecture
to first order in $\alpha-1$ \cite{WieseMajumdarRosso11}. This perturbative approach was later employed
to compute not just exponents but complete scaling scaling functions of various FBM observables
\cite{DelormeWiese15,DelormeWiese16,ArutkinWalterWiese20}.

In the present paper, we focus on comparing the probability densities of the FLE and FBM in the presence of absorbing walls.
We consider two geometries, (i) a semi-infinite interval having an absorbing wall at the origin and (ii)
a finite interval with absorbing boundaries at both ends and a source in the center. Our paper is organized as
follows. FBM and the FLE are introduced in Secs.\ \ref{sec:FBM} and \ref{sec:FLE}, respectively, where we
also outline our numerical approaches. For comparison purposes, Sec.\ \ref{sec:normal} summarizes
the behavior of normal diffusion with absorbing walls.  The simulation results for FBM with absorbing walls
are given in Sec.\ \ref{sec:results_FBM}. Section \ref{sec:results_FLE} presents the corresponding results for
the FLE together with  comparison between the two stochastic processes. We conclude in Sec.\ \ref{sec:conclusions}.

%%%%%%%%%%%%%%%%%%%%%%%%%%%%%%%%%%%%%%%%%%%%%%%%%%%%%%%%%%%%%%%%%%%%%%%
\section{Fractional Brownian motion}
\label{sec:FBM}
%%%%%%%%%%%%%%%%%%%%%%%%%%%%%%%%%%%%%%%%%%%%%%%%%%%%%%%%%%%%%%%%%%%%%%%
\subsection{Definition}
\label{subsec:Def_FBM}
%%%%%%%%%%%%%%%%%%%%%%%%%%%%%%%%%%%%%%%%%%%%%%%%%%%%%%%%%%%%%%%%%%%%%%%

Unconfined FBM is a non-Markovian continuous-time Gaussian stochastic process with stationary
increments. It is centered, i.e., the average
of the position $X$ at time $t$ vanishes, $\langle X(t)\rangle=0$.
The position covariance function reads
\begin{equation}
\langle X(s) X(t) \rangle = K (s^\alpha - |s-t|^\alpha + t^\alpha)~
\label{eq:FBM_cov}
\end{equation}
The exponent $\alpha$ can take values in the range $0 < \alpha < 2$.
The mean-square displacement of the process is obtained from
eq.\ (\ref{eq:FBM_cov}) by setting $s=t$, resulting in anomalous diffusion
$\langle X^2 \rangle = 2 K t^\alpha$, i.e., superdiffusion for $\alpha>1$
and subdiffusion for $\alpha<1$. The marginal case $\alpha=1$ corresponds to
normal diffusion. The increment process $\xi(t)$ defined via
\begin{equation}
X(t) = \int_0^t dt' \xi(t')
\label{eq:FGN_continuous}
\end{equation}
constitutes a fractional Gaussian noise \cite{Qian03}, i.e., a stationary Gaussian process of zero mean and covariance
\begin{equation}
\langle \xi(t) \xi(t')\rangle = K \alpha (\alpha-1) |t-t'|^{\alpha-2}     \qquad (t\ne t')~.
\label{eq:FGN_cov_continuous}
\end{equation}
The correlations are positive (persistent) for $\alpha>1$ and negative (anti-persistent)
for $\alpha < 1$. In the marginal case, $\alpha=1$, the covariance vanishes for all
$t\ne t'$. The probability density of unconfined FBM takes the Gaussian form
\begin{equation}
	P(x,t) = \frac{1}{\sqrt{4\pi K t^\alpha}} \exp{ \left( -\frac{x^2}{4 K t^\alpha} \right) }~.
\label{eq:FBM_P(x)_free}
\end{equation}

In preparation for the computer simulations, we now discretize time using a time step $\epsilon$
by setting $t_n= \epsilon n$ (where $n$ is an integer) and $x_n = X(t_n)$. The resulting discrete version
of FBM \cite{Qian03} can be understood as a random walk with identically Gaussian distributed but
long-time correlated increments. Specifically, the position $x_n$ of the particle follows the recursion relation
\begin{equation}
x_{n+1} = x_n + \xi_n~.
\label{eq:FBM_recursion}
\end{equation}
The increments $\xi_n$ are Gaussian random numbers of zero mean, variance $\sigma^2 = 2 K \epsilon^\alpha$,
and covariance function
\begin{equation}
C_n=\langle \xi_m \xi_{m+n} \rangle = \frac 1 2 \sigma^2 (|n+1|^\alpha - 2|n|^\alpha + |n-1|^\alpha)~.
\label{eq:FGN_cov}
\end{equation}
 In the long-time limit
$n\to \infty$, this covariance takes the power-law form $\langle \xi_m \xi_{m+n} \rangle  \sim\alpha (\alpha-1) |n|^{\alpha-2}$,
in agreement with the continuum version (\ref{eq:FGN_cov_continuous}).

To approximate the continuum limit, the time step $\epsilon$ of the discrete FBM must be much smaller than
the considered total times $t$. Equivalently, the standard deviation $\sigma$ of an individual increment
must be small compared to typical distances or system sizes.
This can be achieved either by taking $\epsilon$ to zero at fixed $t$ or, equivalently,
by taking $t$ to infinity at fixed $\epsilon$.

Absorbing walls can be easily implemented by suitably modifying the recursion relation
(\ref{eq:FBM_recursion}). As the fractional Gaussian noise is understood as externally given
\cite{Klimontovich_book95}, it is not affected by the walls. Specifically, an absorbing
wall at position $w$ that confines the particles to $x \ge w$ (i.e., a wall to the left
of the allowed interval) can be defined by simply removing the particle whenever $x_n < w$.
A wall confining the motion to positions $x \le w$ (i.e., a wall at the right end of an
allowed interval) can be defined analogously. We note in passing that the definition of
\emph{reflecting} walls is not as straight forward because there is some ambiguity in where to place
the particle after the interaction with the wall. For a discussion of this issue see, e.g.,
Ref.\ \cite{VHSJGM20} and references therein.

%%%%%%%%%%%%%%%%%%%%%%%%%%%%%%%%%%%%%%%%%%%%%%%%%%%%%%%%%%%%%%%%%%%%%%%
\subsection{Simulation details}
\label{subsec:Simulations_FBM}
%%%%%%%%%%%%%%%%%%%%%%%%%%%%%%%%%%%%%%%%%%%%%%%%%%%%%%%%%%%%%%%%%%%%%%%

We perform extensive computer simulations of the discrete-time FBM defined by eqs.\
(\ref{eq:FBM_recursion}) and (\ref{eq:FGN_cov}) for several anomalous diffusion exponents $\alpha$
between 0.6 (subdiffusive regime) and 1.6 (superdiffusive regime).
We set the time step to $\epsilon=1$ and fix $K$ at $K=1/2$ . The resulting standard deviation
of the individual increments $\xi_n$ is $\sigma=1$.

We consider two geometries. In the first set of simulations, the particles start at position
$x_0$ and are confined to nonnegative $x$ values by an absorbing wall at the origin, $x=0$.
The maximum time ranges from $2^{17} \approx 130,000$ deep in the superdiffusive regime ($\alpha=1.6$)
to $2^{23} \approx 8.4$ million for the most subdiffusive $\alpha=0.6$. These long times allow
us to reach the continuum (scaling) limit for which the time discretization becomes unimportant.
Because a significant fraction of the particles is absorbed by the absorbing wall before the
final time, these simulations require large numbers of particles.
We start the simulations with $10^8$ to $10^9$ particles to reach a statistical accuracy that
allows us to analyze the probability density at the final time.

In the second set of simulations, the particles start at the origin and are confined to the
finite interval $[-L/2, L/2]$ by absorbing walls at both ends. If a particle is absorbed by one
of the walls, it is not removed but placed back at the origin (and the memory of its previous
steps is erased). This models a particle source at the center of the interval. In these
simulations, we focus on the steady state that is reached after sufficiently long times.
It is characterized by a constant particle current from the center of the interval to the absorbing
walls. We employ interval length ranging from $L=4\times 10^5$ for $\alpha=1.6$ to $L=400$ for
$\alpha=0.6$. All lengths fulfill the condition $L/\sigma \gg 1$ necessary to reach the
continuum (scaling) limit for the discrete FBM.
As no particles are lost in these simulations, they require smaller particle numbers. We use between
$10^4$ and $5 \times 10^4$ particles for each simulation run. The longest times reached for each
$\alpha$ are in the
range from $2^{25} \approx 33$ million to $2^{27} \approx 134$ million.

The correlated Gaussian random numbers $\xi_n$  that represent the discrete fractional Gaussian noise
(the FBM increments or steps)
are precalculated before each particle performs the random walk. (If a particle is placed back at the
origin after absorption in the second set of simulations, its set of random numbers is discarded, and
a new set of random numbers is created.)
The random numbers are created by means of the Fourier-filtering technique \cite{MHSS96} which
consists of the following steps: First, a sequence of independent
Gaussian random numbers $\chi_i$ of zero average and unit variance is created
(using the Box-Muller transformation with the LFSR113 random number generator
proposed by L'Ecuyer \cite{Lecuyer99} as well as
the 2005 version of Marsaglia's KISS \cite{Marsaglia05}).
The discrete Fourier transform $\tilde \chi_\omega$ of these numbers is then converted via
${\tilde{\xi}_\omega} = [\tilde C(\omega)]^{1/2} \tilde{\chi}_\omega$,
where $\tilde C(\omega)$ is the Fourier transform of the covariance function (\ref{eq:FGN_cov})
of the fractional Gaussian noise. The desired correlated noise values $\xi_n$ are given by
the inverse Fourier transformation of the  ${\tilde{\xi}_\omega}$.

The computational effort of the Fourier-filtering technique scales as $N_t \ln N_t$ with the number of
time steps while the effort for the actual propagation of the recursion relation (\ref{eq:FBM_recursion})
scales linearly with $N_t$. Overall, our simulations thus show a favorable,  almost linear scaling
with $N_t$. This allows us to reach long times and employ large numbers of particles.

%%%%%%%%%%%%%%%%%%%%%%%%%%%%%%%%%%%%%%%%%%%%%%%%%%%%%%%%%%%%%%%%%%%%%%%
\section{Fractional Langevin equation}
\label{sec:FLE}
%%%%%%%%%%%%%%%%%%%%%%%%%%%%%%%%%%%%%%%%%%%%%%%%%%%%%%%%%%%%%%%%%%%%%%%
\subsection{Definition}
\label{subsec:Def_FLE}
%%%%%%%%%%%%%%%%%%%%%%%%%%%%%%%%%%%%%%%%%%%%%%%%%%%%%%%%%%%%%%%%%%%%%%%

The well-known normal Langevin equation \cite{Langevin08},
\begin{equation}
m \frac {d^2}{dt^2} x(t) = - \bar\gamma \frac d {dt} x(t) + \xi_w(t)~,
\label{eq:LE}
\end{equation}
describes the motion of a particle of mass $m$ under
the influence of an uncorrelated random force (Gaussian white noise) $\xi_w(t)$
and a linear instantaneous damping force with damping coefficient $\bar\gamma$.

The Langevin equation can be generalized  \cite{Zwanzig_book01,Haenggi78,Goychuk12}
by considering a correlated random force $\xi(t)$ and a nonlocal (in time) damping force,
leading to the equation
\begin{equation}
m \frac {d^2}{dt^2} x(t) = - \bar\gamma \int_{0}^t dt' \mathcal K(t-t') \frac d {dt'} x(t') + \xi(t)~.
\label{eq:GLE}
\end{equation}
If the noise covariance and the damping (memory) kernel $\mathcal K$ fulfill the relation
\begin{equation}
\langle \xi(t) \xi(t') \rangle = k_B T \bar\gamma \mathcal K(t-t')~,
\label{eq:FDT}
\end{equation}
the fluctuation-dissipation theorem \cite{Kubo66} guarantees that the system
reaches thermal equilibrium at temperature $T$ in the long-time limit.
If the random force is a fractional Gaussian noise (as introduced in Sec.\ \ref{subsec:Def_FBM}),
i.e., a stationary Gaussian process of zero mean and covariance (\ref{eq:FGN_cov_continuous}),
the equation is called the fractional Langevin equation (FLE). Note that the exponent $\alpha$
in the fractional Gaussian noise is restricted to the range $1 <\alpha <2$ because the damping
integral in (\ref{eq:GLE}) diverges at $t=t'$ for $\alpha < 1$, and $\alpha > 2$ is unphysical
because it implies correlations that increase with time.

The properties of the unconfined (free-space) FLE are well understood (see, e.g., Ref.\
\cite{MJCB14} and references therein).  If the particle starts from rest at the origin at
time $t=0$, the probability densities of both its velocity and its position are Gaussians
of zero mean. In the long-time limit, the mean-square velocity reaches the equilibrium value $k_B T/m$
required by the equipartition theorem. The mean-square displacement initially shows ballistic
behavior, $\langle x^2(t) \rangle \sim t^{2}$, and then crosses over to anomalous diffusion,
\begin{equation}
\langle x^2(t) \rangle \sim t^{2-\alpha}~,
\label{eq:MSD_free}
\end{equation}
at longer times. This means the anomalous diffusion exponent of the FLE is actually
$2-\alpha$ rather than $\alpha$ as it was for FBM.
It also implies that the FLE with persistent noise, $1 <\alpha <2$, leads to subdiffusion while
FBM with the same noise produces superdiffusion.
This difference between FBM and the FLE stems from the fact that the fluctuation-dissipation
condition (\ref{eq:FDT}) combines persistent noise with long-time memory in the damping force.

Absorbing walls can be introduced in complete analogy to FBM in Sec.\ \ref{subsec:Def_FBM}.
An absorbing wall at position $w$ that confines the particles to $x \ge w$ is defined by simply
removing the particle whenever $x(t) < w$. A wall confining the motion to positions $x \le w$ is
defined analogously.\footnote{As in the FBM case, \emph{reflecting} walls pose additional challenges;
this has been discussed, e.g., in Ref.\ \cite{VojtaSkinnerMetzler19}.}

%%%%%%%%%%%%%%%%%%%%%%%%%%%%%%%%%%%%%%%%%%%%%%%%%%%%%%%%%%%%%%%%%%%%%%%
\subsection{Simulation method}
\label{subsec:Simulation_FLE}
%%%%%%%%%%%%%%%%%%%%%%%%%%%%%%%%%%%%%%%%%%%%%%%%%%%%%%%%%%%%%%%%%%%%%%%

Our computer simulations of the FLE follow the efficient approach detailed in Ref.\ \cite{VojtaSkinnerMetzler19}.
In the present section, we provide a brief outline of the method.

We start by discretizing the time variable, $t_n= \epsilon n$ (where $n$ is an integer) and
replacing the time derivatives in the FLE by first-order finite-difference expressions.
This yields the recursion relations
\begin{eqnarray}
v_{n+1} &=& v_n + \epsilon \left[\xi_n  - \sum_{m=0}^n {\mathcal K}_{n-m} v_m \right]~,
\label{eq:FLE_discrete_v}
\\
x_{n+1} &=& x_n + \epsilon v_n
\label{eq:FLE_discrete_x}
\end{eqnarray}
which are easily propagated numerically.
The $\xi_n$ are a discrete fractional Gaussian noise \cite{Qian03}, i.e.,
identical Gaussian random numbers of zero mean and covariance
(\ref{eq:FGN_cov}), and we have fixed the mass $m$, the damping coefficient
$\bar\gamma$, and the Boltzmann constant $k_B$ at unity.
The damping kernel fulfills the discrete version of the fluctuation-dissipation theorem,
\begin{equation}
T {\mathcal K}_n = \langle \xi_m \xi_{m+n} \rangle ~.
\label{eq:FDT_discrete}
\end{equation}

The numerical effort for creating the correlated random forces by means of the Fourier filtering
method scales as $N_t \ln N_t$ with the number of time steps,
as explained in Sec.\ \ref{subsec:Simulations_FBM}. However, a naive
implementation of the recursion relation (\ref{eq:FLE_discrete_v}) leads to an unfavorable
quadratic scaling of the effort with $N_t$ because the damping sum contains $O(N_t)$ terms for
every single time step.  We have therefore developed an improved algorithm that speeds up the evaluation
of the damping integrals by several several orders of magnitude. It is based on the fact that the
damping kernel is small and slowly varying for large time lag $n-m$ and effectively reduces the
scaling of the numerical effort from $N_t^2$ to $N_t^{1.2}$, albeit with a
large prefactor. This algorithm is
discussed in detail in the Appendix of Ref.\ \cite{VojtaSkinnerMetzler19}.

\begin{figure}
\centerline{\includegraphics[width=12cm]{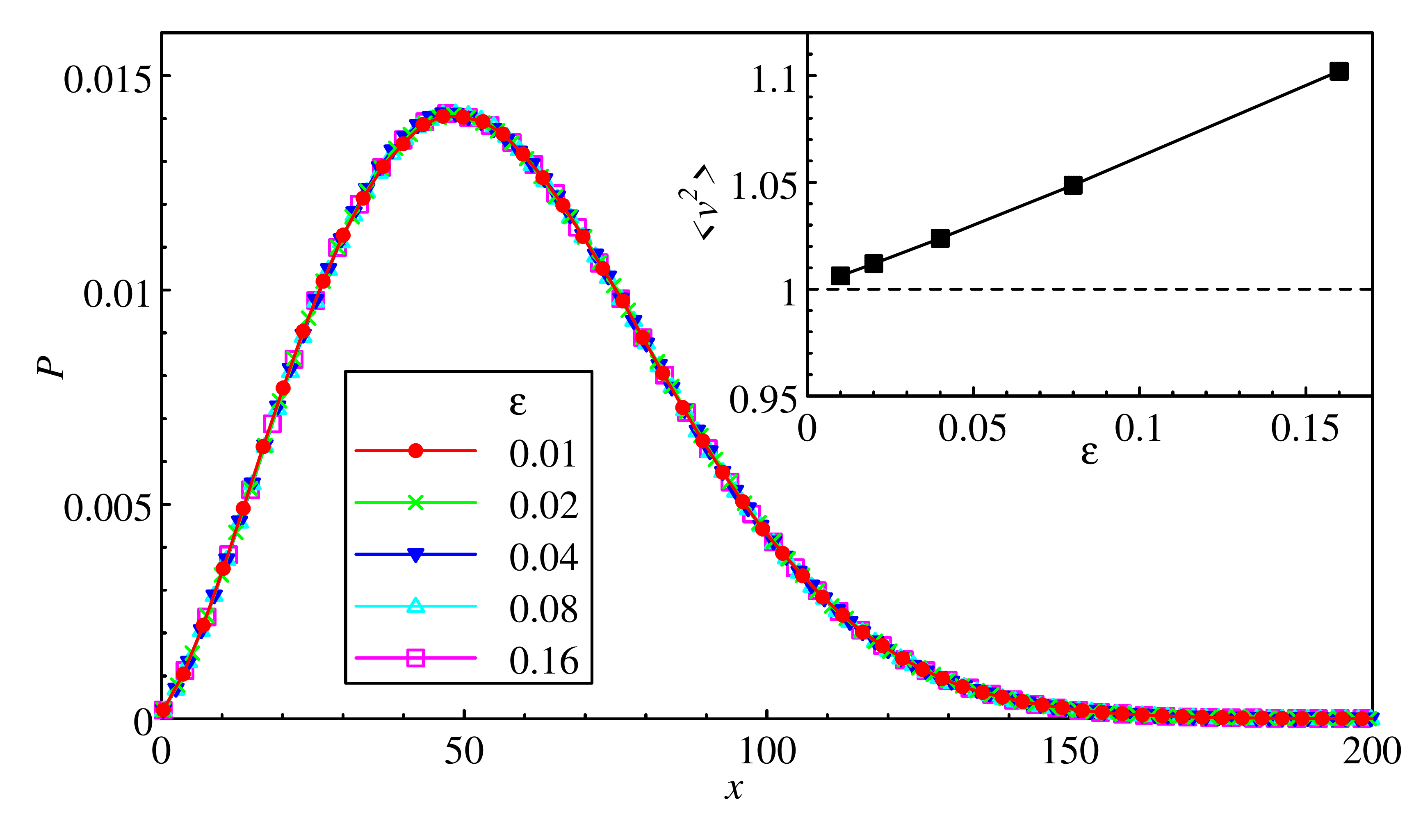}}
\caption{Determination of a suitable value of the time step $\epsilon$.
Main panel: probability density $P(x,t)$ of the particle position at time $t=5234$ for correlation exponent $\alpha=1.2$, temperature
$T=1$, and several values of the time step $\epsilon$. The particles start $x_0=10$ at time $t=0$ and move on the semi-infinite
interval $(0,\infty)$ with an absorbing wall at the origin.
Inset: Stationary mean-square velocity $\langle v^2 \rangle$ as a function of time step $\epsilon$
(the data are averages of $\langle v^2 \rangle$ over the surviving particles during the time intervals $1000 - 5000$).
The statistical error is smaller than the symbol size.
 }
\label{fig:v2vsdt}
\end{figure}

The performance of the simulations depends on choosing a suitable value for the time step
$\epsilon$.  On the one hand, $\epsilon$ should be small to reduce the time discretization
error.  On the other hand, a small $\epsilon$ increases the numerical effort to reach long times.
To optimize the time step, we study how the stationary value of the mean-square velocity
$\langle v^2 \rangle$ depends on the time step. Results for particles moving on the semi-infinite
interval $[0,\infty)$ are shown in the inset of of Fig.\ \ref{fig:v2vsdt}.
The data indicate that the deviation of $\langle v^2 \rangle$ from the value of 1 (required by the
fluctuation-dissipation theorem for temperature $T=1$) decreases with decreasing $\epsilon$, as expected.
The relative error is below 5\% for times steps $\epsilon=0.08$ and smaller.
We also test how the
probability density $P(x,t)$ is affected by the time step $\epsilon$. The main panel of Fig.\
\ref{fig:v2vsdt} presents results for $\alpha=1.2$ and $t=5243$. They show that the probability
densities for time steps between $\Delta t=0.01$ and 0.16 fall right on top of each other.
In the majority of our simulations, we therefore employ time steps $\epsilon =0.04$ and 0.08.

All simulations are performed for noise amplitude $K=1$ and temperature $T=1$, using
the same two geometries as for FBM. In a first set of calculations, particles start at a position $x_0>0$ and are confined
to the nonnegative $x$ axis by an absorbing wall at $x=0$. The longest simulations use $2^{25}$ time steps
of size $\epsilon=0.08$ giving a maximum time of about $2.7 \times 10^6$. The initial particle number is
between $10^7$ and $5 \times 10^8$. In the second set of calculations, particles start at the origin and are confined to the
interval $[-L/2, L/2]$ by absorbing walls at both ends. When a particle is absorbed by one of the walls,
it is placed back at the origin. We consider two restart conditions, (i) the particle retains its random force sequence
and velocity memory or (ii) the particle restarts with a fresh set of random
forces and has no memory of its velocities before the absorption. Condition (ii) is perhaps more physically
plausible and corresponds to the method used for FBM. Most of our simulations thus employ condition (ii),
but we also carry out a few test calculations using condition (i). The two cases are expected to produce
slightly different steady states, but with the same qualitative behavior close to the absorbing walls.
We use up to $2^{29}$ time steps and employ interval length between 400 and $10^5$, depending on $\alpha$.
As no particles are lost in these simulations, they require smaller particle numbers. We use between
$5000$ and $20000$ particles for each run.

%%%%%%%%%%%%%%%%%%%%%%%%%%%%%%%%%%%%%%%%%%%%%%%%%%%%%%%%%%%%%%%%%%%%%%%
\subsection{Antipersistent noise}
\label{subsec:FLE_antipersistent}
%%%%%%%%%%%%%%%%%%%%%%%%%%%%%%%%%%%%%%%%%%%%%%%%%%%%%%%%%%%%%%%%%%%%%%%

In the continuous-time FLE (\ref{eq:GLE}), the exponent $\alpha$ of the random force
is restricted to $1 < \alpha <2$, i.e., to the case of persistent noise.
Anti-persistent noise, $\alpha<1$ is impossible because the damping integral diverges
in this case. Because the divergence stems from short time lags near $t=t'$, it can be
cut off without modifying the physically important long-time behavior of the the
damping kernel. In the discretized FLE defined via eqs.\ (\ref{eq:FLE_discrete_v}) and
(\ref{eq:FLE_discrete_x}) the singularity is already cut off as the damping sum
in (\ref{eq:FLE_discrete_v}) remains finite for all correlation exponents in the
full interval $0 < \alpha < 2$. This allows us to extend our simulation to the
anti-persistent case.

The behavior of the discretized FLE in the regime $\alpha  < 1$ is rather peculiar.
The (discrete) fluctuation-dissipation theorem (\ref{eq:FDT_discrete}) implies
that the damping kernel ${\mathcal K}_{n-m}$ has negative values for $n \ne m$ if the
noise is antipersistent. Instead of damping, these terms thus yield antidamping,
i.e., a positive feedback for the velocity. Consequently, the mean-square
displacement for the free, unconfined FLE, $\langle x^2 \rangle \sim t^{2-\alpha}$,  grows super-diffusively
for $\alpha < 1$.

%%%%%%%%%%%%%%%%%%%%%%%%%%%%%%%%%%%%%%%%%%%%%%%%%%%%%%%%%%%%%%%%%%%%%%%
\section{Normal diffusion with absorbing walls}
\label{sec:normal}
%%%%%%%%%%%%%%%%%%%%%%%%%%%%%%%%%%%%%%%%%%%%%%%%%%%%%%%%%%%%%%%%%%%%%%%

For comparison purposes, this section briefly summarizes the relevant results for normal diffusion.
In this case, the probability density $P(x,t)$  can be
found by solving the diffusion equation $\partial_t P(x,t) = K \partial_{xx} P(x,t)$.

In our first geometry, all particles start at $x_0>0$ at time $t=0$, and are confined to the non-negative $x$-axis
by an absorbing wall at the origin, $x=0$. The diffusion equation thus has the
initial condition $P(x,0) = \delta(x-x_0)$ and the absorbing
boundary condition $P(0,t)=0$. This problem can be solved easily by means of the method of
images \cite{Redner_book01}, giving
\begin{equation}
P(x,t) = \frac {1} {\sqrt{4\pi K t}} \left[ \exp{ \left( -\frac{(x-x_0)^2}{4 K t} \right)}
-\exp{ \left( -\frac{(x+x_0)^2}{4 K t} \right) } \right]~.
\label{eq:P_images}
\end{equation}
Integrating $P(x,t)$ over all $x \ge 0$ yields the survival probability
\begin{equation}
S(t)= \frac {1}{\sqrt{4 \pi K t}} \int_{-x_0}^{x_0} dx \exp{ \left( -\frac{x^2}{4 K t} \right)} \to
  \begin{cases}
   1                  & \text{~if~ } 2Kt  \ll x_0^2  \\
   x_0/\sqrt{\pi K t} & \text{~if~ } 2Kt \gg x_0^2
  \end{cases} ~.
\label{eq:S_images}
\end{equation}
The conditional probability density $P_s(x,t)= P(x,t)/S(t)$ for the surviving particles takes
the form
\begin{equation}
P_s(x,t) = \frac x {2Kt} \exp{ \left( -\frac{x^2}{4 K t} \right)}
\label{eq:Ps_images}
\end{equation}
in the long time limit $2Kt  \gg x_0^2$. The corresponding mean square displacement of the surviving particles
from the origin reads $\langle x^2 \rangle = 4 Kt$ (twice the value of free Brownian motion).

In our second geometry, we consider the steady state on a finite interval $[-L/2,L/2]$ with absorbing walls
at both ends and a source in the center. The stationary solution of the diffusion equation with
boundary conditions $P(-L/2,t)=P(L/2,t)=0$ reads
\begin{equation}
P(x,t) = \frac 2 L \left ( 1-\frac {2|x|} L \right ) ~,
\label{eq:P_stationary_normal}
\end{equation}
when it is properly normalized to unity.

%%%%%%%%%%%%%%%%%%%%%%%%%%%%%%%%%%%%%%%%%%%%%%%%%%%%%%%%%%%%%%%%%%%%%%%
\section{Simulation results for FBM with absorbing walls}
\label{sec:results_FBM}
%%%%%%%%%%%%%%%%%%%%%%%%%%%%%%%%%%%%%%%%%%%%%%%%%%%%%%%%%%%%%%%%%%%%%%%
\subsection{Semi-infinite interval}
\label{subsec:semi-infinite_FBM}
%%%%%%%%%%%%%%%%%%%%%%%%%%%%%%%%%%%%%%%%%%%%%%%%%%%%%%%%%%%%%%%%%%%%%%%

\begin{figure}
\centerline{\includegraphics[width=16cm]{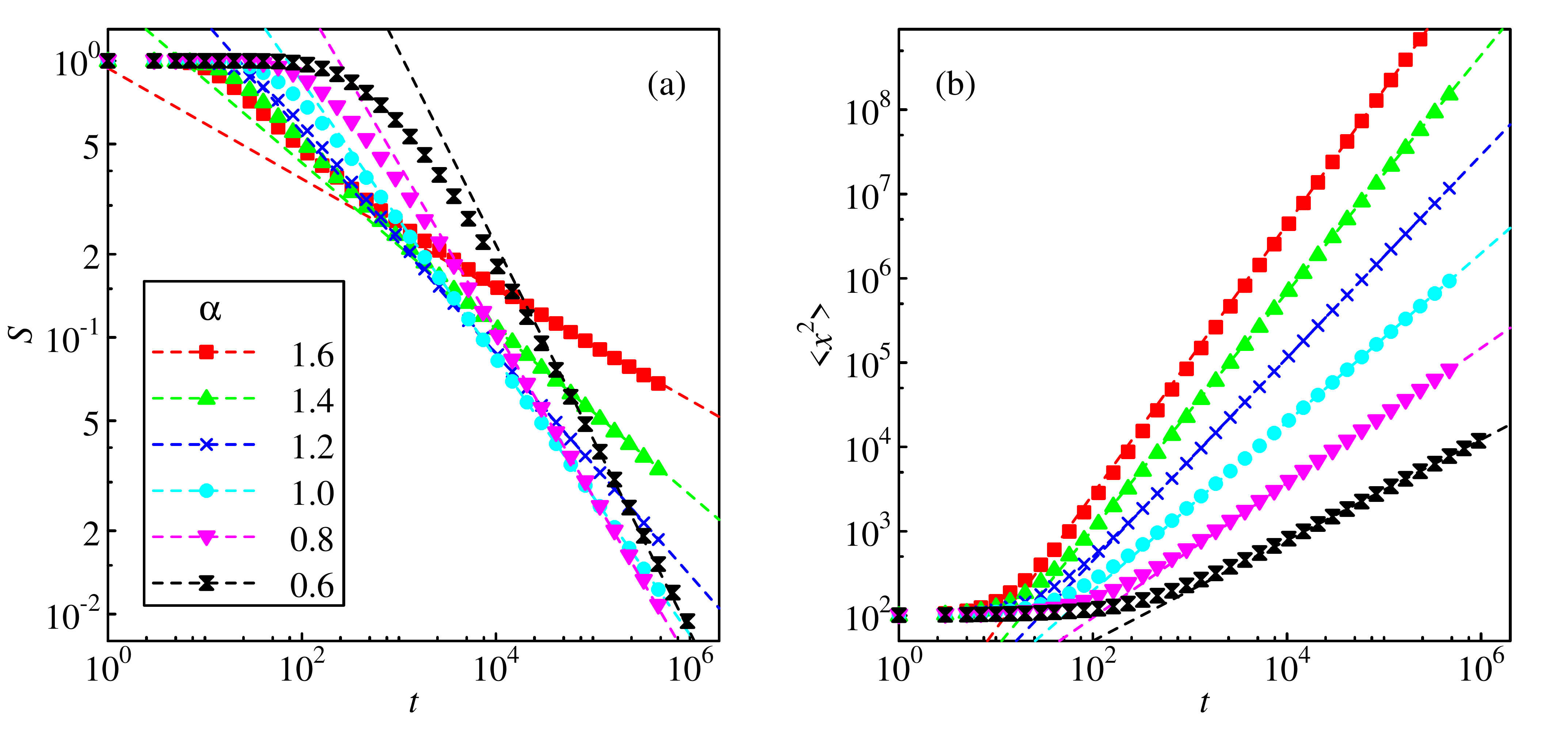}}
\caption{FBM on the semi-infinite interval $[0,\infty)$. (a) Survival probability $S$ vs.\ time $t$ for several values of $\alpha$.
The initial particle number ranges from $10^7$ to $10^8$, all particles start at $x_0=10$.
The dashed lines are fits of the long-time behavior to the expected
power law $S(t) \sim t^{-(1-\alpha/2)}$.
(b) Mean square displacement from the origin $\langle x^2 \rangle$ of the surviving particles vs.\ time $t$
for the same simulations as in panel (a).
The dashed lines are fits of the long-time behavior to the expected
power law $\langle x^2 \rangle \sim t^\alpha$.
The statistical errors of all data points are smaller than the symbol size.}
\label{fig:fbm_nn_x2_all}
\end{figure}

In this section, we report the results of FBM simulations in which the particles
start at a position $x_0 > 0$ at time $t=0$ and are confined to
the semi-infinite interval $[0,\infty)$ by an absorbing wall at the origin.
Figure \ref{fig:fbm_nn_x2_all}
gives an overview over the time evolution of the
survival probability $S(t)$ and mean-square displacement $\langle x^2 \rangle$ from the origin
of the surviving particles for $x_0=10$ and several values of the anomalous diffusion
exponent $\alpha$ that cover the range from subdiffusive to superdiffusive
behavior.
The long-time decay of the survival probability is expected to follow the
power-law $S(t) \sim t^{-\theta}$ with $\theta$ being the persistence exponent,
which is known exactly for FBM, $\theta=1-\alpha/2$ \cite{KKMCBS97,ZoiaRossoMajumdar09}.
The data in Fig.\ \ref{fig:fbm_nn_x2_all}(a) agree very well with this power law for all
studied values of $\alpha$. The mean-square displacement from the origin
for the surviving particles, shown in Fig.\ \ref{fig:fbm_nn_x2_all}(b) follows
the same anomalous diffusion power law, $\langle x^2 \rangle \sim t^\alpha$, as it would
for free FBM. We also note that, for normal Brownian motion $\alpha=1$, the prefactors of both
 power laws agree with the diffusion
equation results of Sec.\ \ref{sec:normal} with high accuracy.
Simulations starting from different initial positions $x_0$ lead to analogous results.
In fact, the survival probability curves for different $x_0$ collapse onto each other
when plotted as a function of $t/x_0^{2/\alpha}$ implying the scaling from
$S(t,x_0)= F(t/x_0^{2/\alpha})$. The mean-square displacement of the \emph{surviving} particles
becomes independent of $x_0$ for sufficiently long times.

We now turn to the time evolution of the probability density $P(x,t)$, starting from the initial
condition $P(x,0) = \delta(x-x_0)$. We mostly discuss the conditional probability
density $P_s(x,t)= P(x,t)/S(t)$ for the surviving particles, which is properly normalized
to unity for all times.

\begin{figure}
\centerline{\includegraphics[width=16cm]{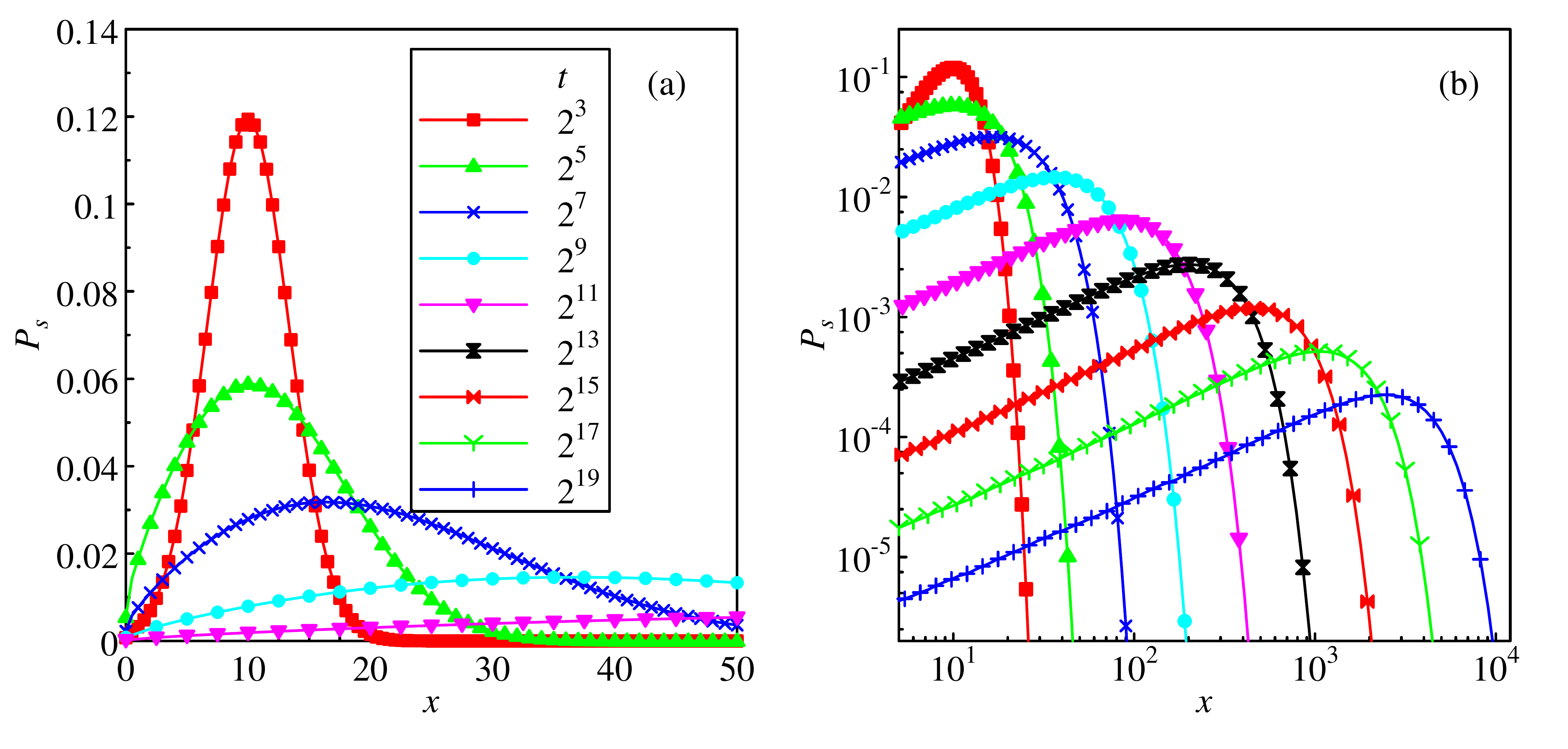}}
\caption{Time evolution of the conditional probability density $P_s(x,t)$ for FBM with anomalous diffusion
exponent $\alpha=1.2$, confined to the positive $x$-axis by an absorbing wall at the origin.
The initial particle number is $5 \times 10^8$, all particles start at $x_0=10$. To further improve the
statistics, $P_s$ is averaged over a short time interval around the given time.
(a) Linear plot of $P_s$ vs.\ $x$ covering the early time behavior.
(b) Double-logarithmic plot of $P_S$ vs.\ $x$ for all simulated times.
The statistical errors of all data points are smaller than the symbol size.}
\label{fig:distrib_08_all_times}
\end{figure}

Let us start by focusing on the value $\alpha=1.2$ of the anomalous diffusion exponent.
Figure \ref{fig:distrib_08_all_times} illustrates the time evolution of $P_s(x,t)$ in this case.
Panel \ref{fig:distrib_08_all_times}(a) shows that $P_s(x,t)$ follows, for early times,
a Gaussian distribution centered at $x_0=10$ whose width grows like
$t^{\alpha/2}$, just as in the case of free, unconfined FBM.
Once the distribution starts interacting with the absorbing wall, the functional form of
$P_s(x,t)$ changes. Whereas the large-$x$ tail remains Gaussian, the behavior close to the
absorbing wall takes a power-law form with an exponent that is independent of time.
This is demonstrated in Fig.\ \ref{fig:distrib_08_all_times}(b).

\begin{figure}
\centerline{\includegraphics[width=16cm]{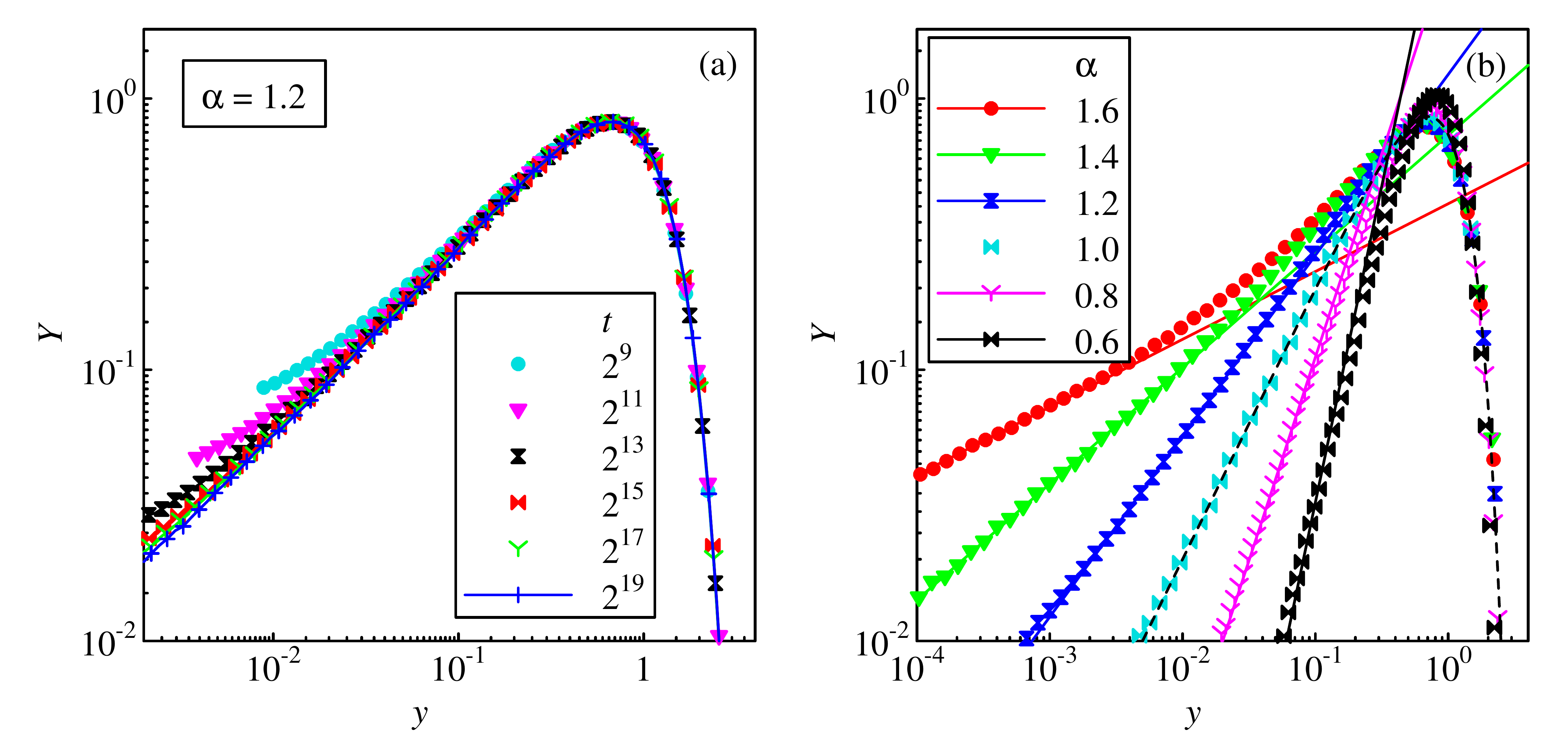}}
\caption{Conditional probability densities of FBM on the semi-infinite interval $[0,\infty)$ for $ x_0=10$,
expressed
in terms of the reduced variables $y=x/x_m(t)$ and $Y = P_s \, x_m(t)$ with $x_m(t)=\langle x^2(t)\rangle^{1/2}$.
(a) Data for $\alpha=1.2$ at different times.
The deviations from the common master curve at the smallest $x$ for each $t$ stem from the time discretization,
they occur for $x$ of the order of the step width $\sigma=1$.
(b) Data for different $\alpha$ at the longest simulated times, ranging from $t=2^{17}$ (for $\alpha=1.6$)
to $t=2^{23}$ (for $\alpha=0.6$). The solid lines are fits of the small-$y$ behavior to the conjectured
power law $Y \sim y^\kappa$ with $\kappa=2/\alpha-1$. The dashed line corresponds to the analytical result
(\ref{eq:Ps_images}) for normal diffusion, without adjustable parameters. The statistical errors of the data points are
about a symbol size (for the smallest shown $Y$) or smaller.
}
\label{fig:distrib_08_all_times_scaled}
\end{figure}

The conditional probability distributions
for sufficiently long times can be collapsed onto a single master curve by plotting the distribution
in terms of the reduced variables $y=x/x_m(t)$ and $Y = P_s \, x_m(t)$
where $x_m(t) = \langle x^2(t)\rangle^{1/2}$ is the root mean square displacement of the surviving
particles from the origin.
Figure \ref{fig:distrib_08_all_times_scaled}(a) indicates a nearly perfect data collapse for
times $t \ge 2^9$, about the time where the survival probability and mean-square
displacement in Fig.\ \ref{fig:fbm_nn_x2_all} reach their asymptotic power-law behaviors.
The figure also shows that the small-$x$ data for each of the curves deviate from the common master curve.
These deviations stem from the time discretization (finite time step $\epsilon$); they occur for $x$ of
the order of the step width $\sigma=1$. As the distribution broadens with increasing time, the onset of
these time discretization artifacts is pushed to smaller $y$ as $t$ increases.
The data collapse in Fig.\ \ref{fig:distrib_08_all_times_scaled}(a) implies that the conditional probability
density fulfills the scaling from
\begin{equation}
P_s(x,t) = \frac 1 {x_m(t)} Y_\alpha^\mathrm{fbm} \left [ \frac x {x_m(t)}\right]
         = \frac 1 {\sqrt{2Kt^{\alpha}}} \tilde Y_\alpha^\mathrm{fbm} \left[ \frac x {\sqrt{2Kt^{\alpha}}} \right ]
\label{eq:Ps_scaling_fbm}
\end{equation}
in the long-time limit  $x_m(t)  \gg x_0$.
Note that the analytical expression (\ref{eq:Ps_images}) for the distribution in the
normal diffusion case, $\alpha=1$, does fulfill this scaling form. Simulations starting
from different initial positions $x_0$ reveal that the conditional probability densities
become independent of $x_0$ for  $x_m(t)  \gg x_0$,
implying that the scaling function $Y_\alpha^\mathrm{fbm}$ does not depend on $x_0$.

We have performed analogous simulations for several $\alpha$ between 0.6 and 1.6. The resulting
conditional probability densities $P_s(x,t)$ all fulfill the scaling form (\ref{eq:Ps_scaling_fbm}). Figure
\ref{fig:distrib_08_all_times_scaled}(b) compares $P_s(x,t)$ for all simulated $\alpha$ (using
the data taken at the longest simulated time for each $\alpha$). The data for the normal diffusion case,
$\alpha=1$, agree nearly perfectly with the analytical solution (\ref{eq:Ps_images}), without adjustable
parameters. This gives us additional confidence in the precision of our simulations.
All distributions show power-law behavior close to the absorbing wall, $y \ll 1$. The data
in this regime can be fitted with high accuracy with the conjectured \cite{ZoiaRossoMajumdar09}
power law $Y \sim y^\kappa$ with $\kappa=2/\alpha-1$. For our most subdiffusive simulation,
$\alpha=0.6$, some slight deviations can be observed between the conjectured power law and
the numerical data (which are slightly less steep). They can be attributed to the simulations
not having progressed far enough into the asymptotic long-time regime.
(The power-law behavior can, at best, be expected in the $x$ range between the step size and
the root mean square displacement, $\sigma \ll x \ll x_m(t)$.
For $\alpha=0.6$, this range is very narrow, $10 \lessapprox x \lessapprox 20$, even at our
longest simulation time $t=2^{23}$.)

\begin{figure}
\centerline{\includegraphics[width=12cm]{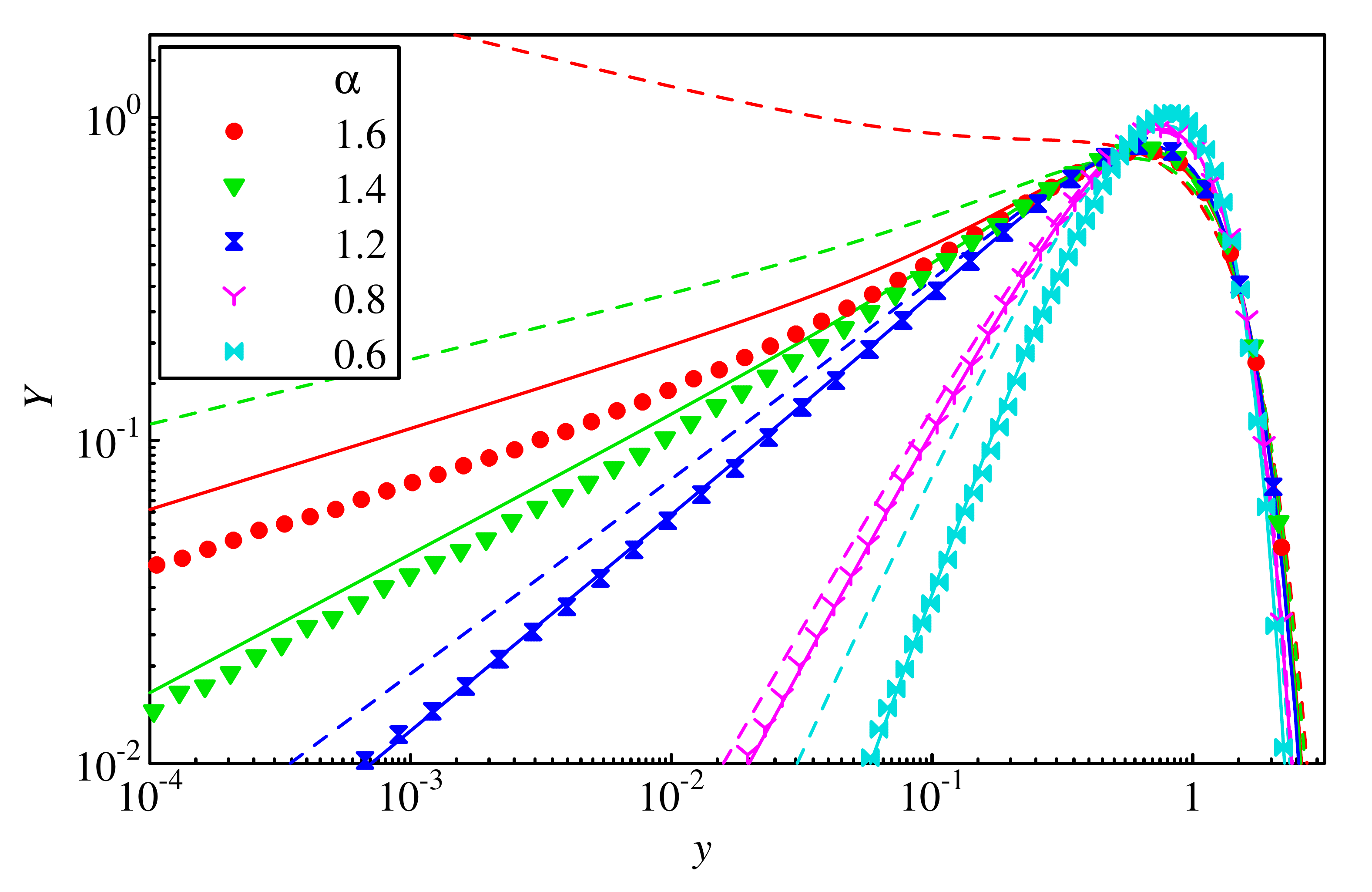}}
\caption{Comparison of the scaled conditional probability densities and the predictions of the
perturbative approach in terms of the scaling variables $y=x/x_m(t)$ and $Y = P_s \, x_m(t)$.
The symbols show our numerical data.
The dashed lines correspond to the first-order expression (\ref{eq:Ps_perturb}) while the solid lines
show the improved result (\ref{eq:Ps_perturb2}). Note that the scaling variables
$y$ and $\tilde y$ are not identical but related by a constant but $\alpha$-dependent factor.
}
\label{fig:perturb_comparison}
\end{figure}

In recent years, a number of analytical results for FBM have been obtained
by means of a perturbative method \cite{WieseMajumdarRosso11,DelormeWiese15,DelormeWiese16,ArutkinWalterWiese20}
in which observables are expanded about the normal Brownian motion case, $\alpha=1$,
with $\epsilon = (\alpha-1)/2$ being the small parameter.
Within this approach, the scaling function $\tilde Y_\alpha^\mathrm{fbm}$ of the conditional probability
density $P_s(x,t)$ [defined in eq.\ (\ref{eq:Ps_scaling_fbm})] can be written in terms of the scaling
variable $\tilde y = x/(2Kt^\alpha)^{1/2}$ as
\begin{equation}
\tilde Y_\alpha^\mathrm{fbm}(\tilde y) = \tilde y\,e^{-\tilde y^2/2} \, e^{\epsilon W(\tilde y)}
\label{eq:Ps_perturb}
\end{equation}
to linear order in $\epsilon$ \cite{DelormeWiese16}. Here, the term $\tilde y\,e^{-\tilde y^2/2}$
represents the solution (\ref{eq:Ps_images}) for normal Brownian motion, while
$e^{\epsilon W(\tilde y)}$ encodes the perturbative correction. The function
$W(\tilde y)$ reads \cite{WieseMajumdarRosso11,DelormeWiese16}
\begin{eqnarray}
W(\tilde y) = &&\frac 1 6 \tilde y^4\,  _2F_2\left(1,1;\frac 5 2, 3; \frac {\tilde y^2} 2\right)
+ \pi (1-\tilde y^2) \mathrm{erfi}\left( \frac {\tilde y} {\sqrt 2} \right)
+ \sqrt{2\pi} e^{\tilde y^2/2} \tilde y \nonumber \\
&&+(\tilde y^2 -2)[\gamma_E + \ln(2 \tilde y^2)] -3 \tilde y^2
\end{eqnarray}
where $_2F_2$ is a hypergeometric function, $\mathrm{erfi}$ is the imaginary error function,
and $\gamma_E$ is the Euler constant. The resulting asymptotic behavior close to the wall (i.e., for $\tilde y \ll 1$)
takes the form $\tilde Y_\alpha^\mathrm{fbm}(\tilde y) \sim \tilde y^{1-4\epsilon} = \tilde y^{3-2\alpha}$.
The exponent $\kappa=1-4\epsilon$ agrees with the conjecture $\kappa=2/\alpha-1$ to linear order in $\epsilon$.
Figure \ref{fig:perturb_comparison} compares our numerical
data for the conditional probability density with the analytical result (\ref{eq:Ps_perturb}).
The figure shows that eq.\ (\ref{eq:Ps_perturb}) describes the bulk of the distributions reasonably
well even for $\alpha$-values far away from $\alpha=1$. However, the power-law behavior close to the 
wall is not reproduced correctly because the first-order approximation of $\kappa$ differs
from the (conjectured) exact value. By combining the conjectured exact power law
$\tilde Y_\alpha^\mathrm{fbm}(\tilde y) \sim \tilde y^{2/\alpha-1}$ with the first-order expression 
(\ref{eq:Ps_perturb}), one can write down an improved approximation \cite{DelormeWiese16},
\begin{equation}
\tilde Y_\alpha^\mathrm{fbm}(\tilde y) = \tilde y^{2/\alpha-1} \,e^{-\tilde y^2/2} \, 
                     e^{\epsilon [W(\tilde y) +4 \ln(\tilde y) +\mathrm{const}] }~. 
\label{eq:Ps_perturb2}
\end{equation}
The expressions (\ref{eq:Ps_perturb}) and (\ref{eq:Ps_perturb2}) agree with each other
to linear order in $\epsilon$ but (\ref{eq:Ps_perturb2}) captures the 
exact asymptotic power law close to the wall. Figure \ref{fig:perturb_comparison}
shows that eq.\ (\ref{eq:Ps_perturb2}) provides a nearly perfect description of the
entire distribution for a sizable $\alpha$ range around $\alpha=1$. Moreover,
it remains a good approximation even for $\alpha$ further away from unity.
 
The simulation of FBM confined to the semi-infinite interval $[0,\infty)$ by an absorbing
wall at the origin is numerically very expensive. One the one hand, long simulation
runs are necessary
to reach the asymptotic long-time regime characterized by $x_m(t) \gg x_0$ and $x_m(t) \gg \sigma$.
On the other hand, the survival probability $S(t)$ drops rapidly with increasing time, requiring
large initial particle numbers in order to keep the statistical error at late times under control.
In the next section, we therefore perform simulations that avoid the particle loss.

%%%%%%%%%%%%%%%%%%%%%%%%%%%%%%%%%%%%%%%%%%%%%%%%%%%%%%%%%%%%%%%%%%%%%%%
\subsection{Stationary state on finite interval}
\label{subsec:finite_FBM}
%%%%%%%%%%%%%%%%%%%%%%%%%%%%%%%%%%%%%%%%%%%%%%%%%%%%%%%%%%%%%%%%%%%%%%%

In this section, we consider FBM on the finite interval $[-L/2,L/2]$ with absorbing walls at both ends.
Particles start at the origin at $t=0$.
When a particle is absorbed by one of the walls, it is placed back at the origin
and the memory of its previous motion is erased. (This is achieved by creating  a new set of
random displacements, completely independent of those before the absorption event.)
This procedure models a particle source at the center of the interval that releases particles at
the same rate as they are absorbed by the walls. The number of particles in these simulations
consequently does not decrease with time. Thus, fewer (initial) particles than in
the semi-infinite interval simulations are sufficient to control the statistical error,
allowing us to reach longer times, up to $t=2^{27}$.

\begin{figure}
\centerline{\includegraphics[width=16cm]{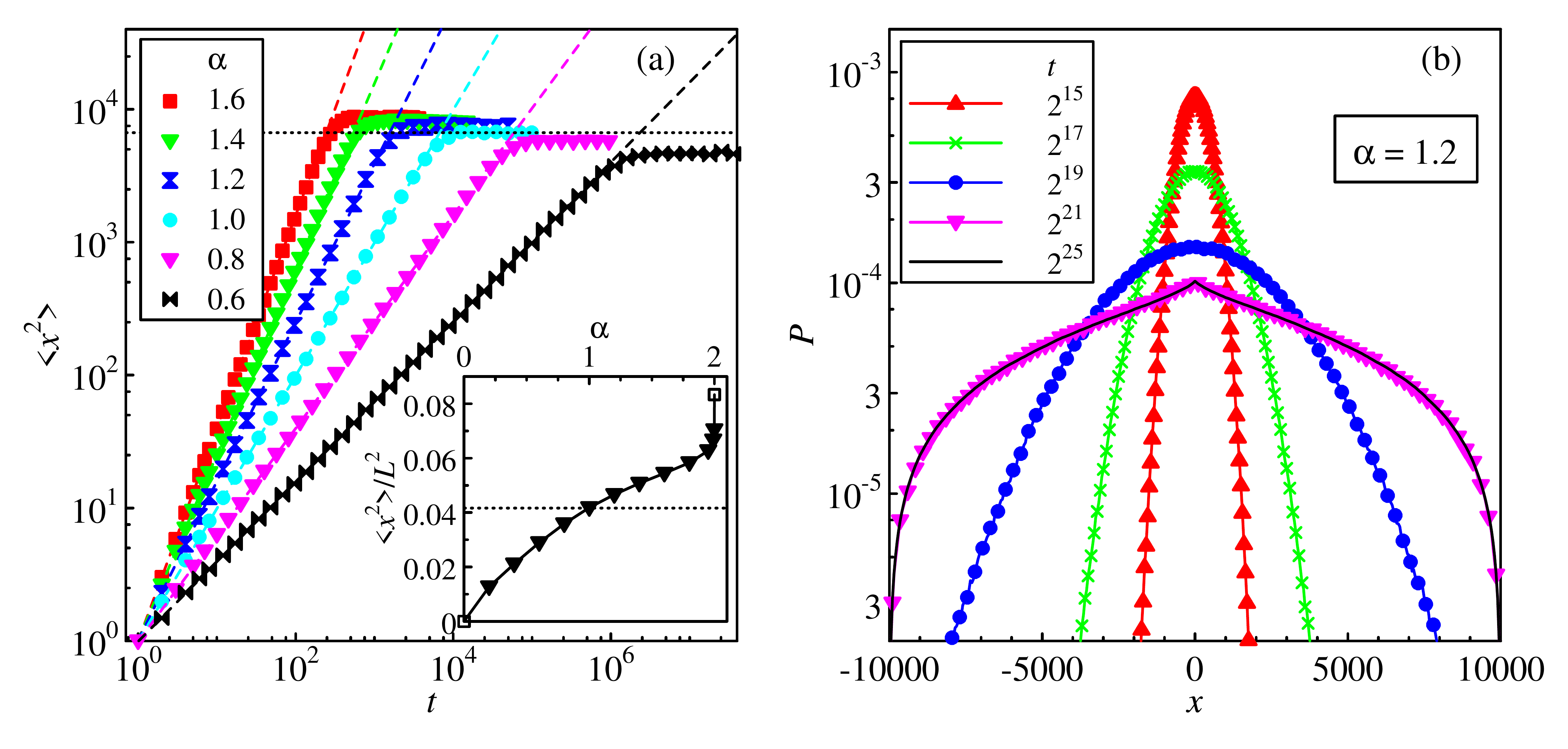}}
\caption{FBM on a finite interval $[-L/2,L/2]$ with absorbing walls at both ends and a source in the center.
(a) Mean square displacement $\langle x^2 \rangle$ vs.\ time $t$ for interval length $L=400$ and several $\alpha$.
The data are averages over 10,000 to 20,000 particles.
The dashed lines are fits to the anomalous diffusion power law $\langle x^2 \rangle \sim t^\alpha$.
The dotted line marks the stationary value $\langle x^2\rangle = L^2/24$ that follows from eq.\
(\ref{eq:P_stationary_normal}) for normal diffusion. Inset: Saturation value of $\langle x^2 \rangle/L^2$ vs.\
$\alpha$. The open squares mark the values $L^2/12$ and 0 expected for $\alpha=2$ and 0, respectively.
(b) Time evolution of the probability density $P(x,t)$ for interval length $L=20,000$ and $\alpha=1.2$
    (between $5\times 10^4$ and $4\times 10^6$ particles used).
    All statistical errors are smaller than the symbol size.
}
\label{fig:fbm_steady_overview}
\end{figure}

Figure \ref{fig:fbm_steady_overview} gives an overview over these simulations.
Panel (a) presents the time dependence of the mean-square displacement for several $\alpha$.
It initially follows the same anomalous diffusion behavior, $\langle x^2 \rangle \sim t^\alpha$
as free, unconfined FBM. Once the particles reach the wall, $\langle x^2 \rangle$ crosses over to
a constant value that varies with $\alpha$, indicating that the steady-state distribution of the
particles itself is $\alpha$ dependent.
In the continuum limit $L \gg \sigma$, the stationary mean-square displacement is proportional
to $L^2$. [This also follows from the scaling law (\ref{eq:P_scaling_interval_fbm}) discussed below.]
The relation between the stationary value of $\langle x^2 \rangle/L^2$
 and $\alpha$ is detailed in the inset of Fig.\ \ref{fig:fbm_steady_overview}(a).
In the normal diffusion case, $\alpha=1$, our result agrees with the value
$\langle x^2\rangle = L^2/24$ that follows from the analytical solution (\ref{eq:P_stationary_normal}).
For $\alpha \to 2$, the stationary value of $\langle x^2 \rangle$ approaches $L^2/12$. This corresponds to the
flat distribution expected for ballistic motion from the source towards the absorbing walls.
For $\alpha \to 0$, the stationary $\langle x^2 \rangle$ approaches zero, as expected.

Figure \ref{fig:fbm_steady_overview}(b) illustrates the time evolution of the probability density for
$\alpha=1.2$. Initially, $P(x,t)$ features the same Gaussian shape as it would for free FBM. Once particles
reach the wall, the shape of $P(x,t)$ changes. For long times (larger than about $2^{21}$ in this example),
$P(x,t)$ become stationary and time-independent.

\begin{figure}
\includegraphics[width=\columnwidth]{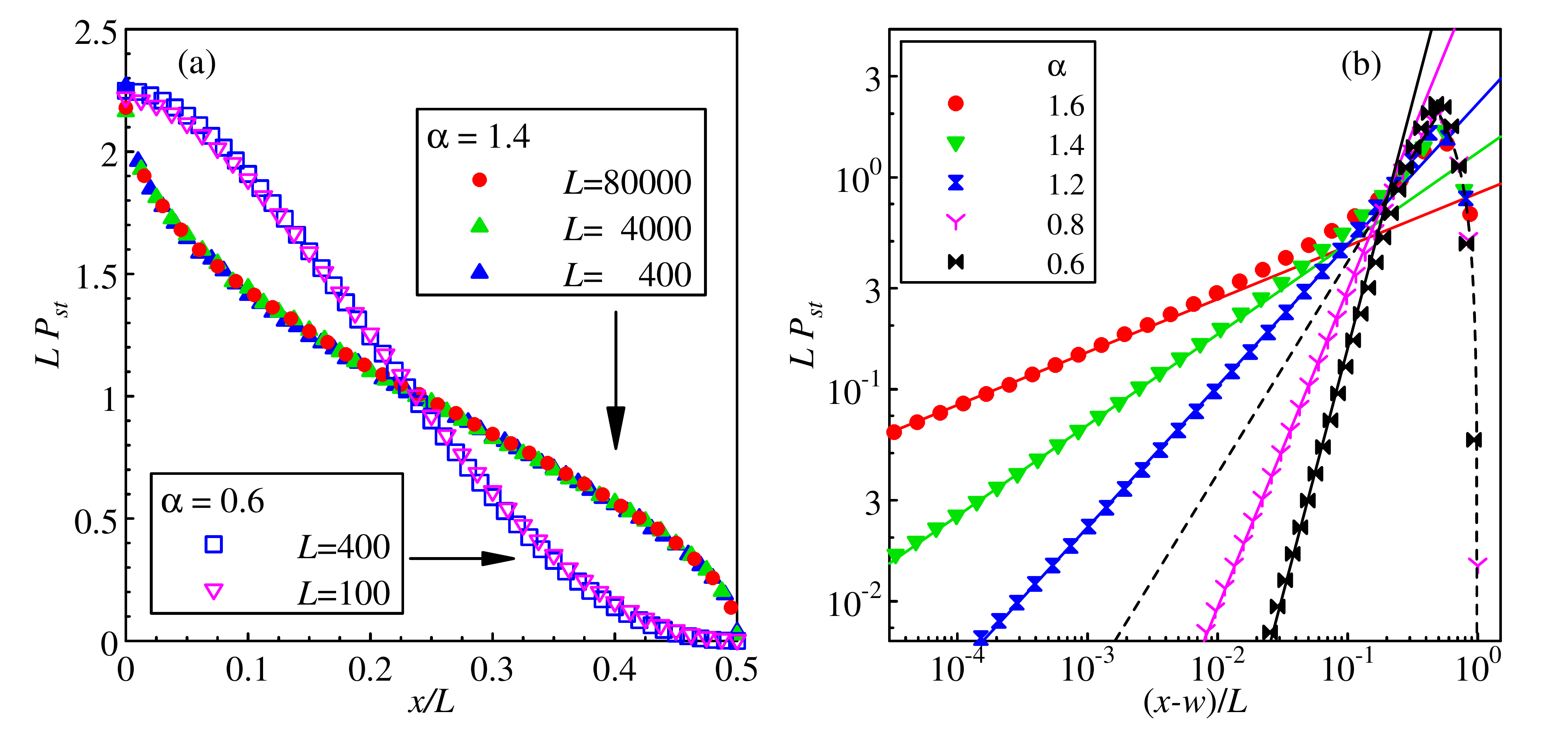}
\caption{(a) Scaling plot of the stationary probability density for FBM on the interval $[-L/2,L/2]$ showing $L P_{st}$ vs.\ $x/L$
for $\alpha=1.4$ and $\alpha=0.6$ for several interval length $L$ (only the right half of the interval is shown).
Each distribution is based
on $10^4$ to $10^5$ particles; $P_{st}$ is averaged over a number of time steps after a stationary state
has been reached. The simulation for $\alpha=0.6$ and $L=400$ required a maximum time of $t=2^{27}$.
(b) Scaled stationary probability density as function of the scaled distance $(x-w)/L$ from the absorbing wall
located at $w=-L/2$ for several
$\alpha$.  The solid lines are fits of the short distance behavior to the
power law $P_{st} \sim (x-w)^\kappa$ with $\kappa=2/\alpha-1$. The dashed line shows the analytical result
(\ref{eq:P_stationary_normal}) for normal diffusion, without adjustable parameters.
The statistical errors of the data are below one symbol size.}
\label{fig:distrib_all_alpha_scaled_interval}
\end{figure}

It is interesting to compare the stationary probability densities $P_{st}$ for several
interval lengths $L$. Figure \ref{fig:distrib_all_alpha_scaled_interval}(a) presents the corresponding simulation data
for several $L$ between 100 and $80000$ using scaled variables $L P_{st}$ vs.\ $x/L$.
The curves for different $L$ collapse nearly perfectly onto a common master curve, demonstrating
that the stationary distribution fulfills the scaling form
\begin{equation}
P_{st}(x,L) = \frac 1 L Z_\alpha^\mathrm{fbm} \left ( \frac x L  \right)~,
\label{eq:P_scaling_interval_fbm}
\end{equation}
where $Z_\alpha^\mathrm{fbm}(z)$ is a dimensionless scaling function.
Small deviations can be attributed to finite-size and finite-time effects
that vanish in the limits $L \gg \sigma$, $2 K t^\alpha \gg L^2$. We have performed analogous
simulations for other values of the anomalous diffusion exponent $\alpha$.
Figure \ref{fig:distrib_all_alpha_scaled_interval}(b) focuses on the behavior of the (scaled) stationary
distribution near the absorbing wall located at $w=-L/2$ for several $\alpha$.
Sufficiently close to the wall, all probability densities follow power laws
in the distance from the wall. They can be fitted with high accuracy
with $P_{st}(x) \sim (x-w)^\kappa$ with $\kappa=2/\alpha-1$,
i.e., with the same exponent as observed in the simulations on the semi-infinite interval.
In addition, we have also analyzed the stationary probability density close to the particle
source in the center of the interval. This analysis is somewhat more complicated
than that near the absorbing walls because the value of probability density at $x=0$
is an additional fitting parameter. Nonetheless, $P_{st}(x)$ for $x \ll L$ can be
fitted well with the expression $P_{st}(x) = a + b |x|^\kappa$ with the same exponent
$\kappa=2/\alpha-1$ for all considered $\alpha$.

Note that FBM on a finite interval with absorbing boundaries at both ends was studied perturbatively 
in Ref.\ \cite{Wiese19}, albeit without the particle source in the center of the interval.

%%%%%%%%%%%%%%%%%%%%%%%%%%%%%%%%%%%%%%%%%%%%%%%%%%%%%%%%%%%%%%%%%%%%%%%
\section{Simulation results for the FLE with absorbing walls}
\label{sec:results_FLE}
%%%%%%%%%%%%%%%%%%%%%%%%%%%%%%%%%%%%%%%%%%%%%%%%%%%%%%%%%%%%%%%%%%%%%%%

Recent simulations have shown that FBM and the FLE behave in qualitatively different
ways in the presence of \emph{reflecting} walls. The probability distribution
of FBM shows strong accumulation (for persistent noise) or depletion (for antipersistent
noise) of particles close to the reflecting wall, both for semi-infinite intervals
\cite{WadaVojta18,WadaWarhoverVojta19} and
for finite intervals \cite{Guggenbergeretal19,VHSJGM20}. In contrast, the fractional Langevin
equation on a finite interval with reflecting walls yields a flat stationary probability density,
just like normal diffusion, for all $\alpha$  \cite{VojtaSkinnerMetzler19}. Some accumulation or
depletion of particles (compared to normal diffusion) occurs on a semi-infinite interval.
However, it is much less pronounced than in the FBM case.
In the present section, we therefore compare and contrast the properties of the FLE with \emph{absorbing}
walls with the those found for FBM in Sec.\ \ref{sec:results_FBM}.

%%%%%%%%%%%%%%%%%%%%%%%%%%%%%%%%%%%%%%%%%%%%%%%%%%%%%%%%%%%%%%%%%%%%%%%
\subsection{Semi-infinite interval}
\label{subsec:semi-infinite_FLE}
%%%%%%%%%%%%%%%%%%%%%%%%%%%%%%%%%%%%%%%%%%%%%%%%%%%%%%%%%%%%%%%%%%%%%%%

\begin{figure}
\centerline{\includegraphics[width=16cm]{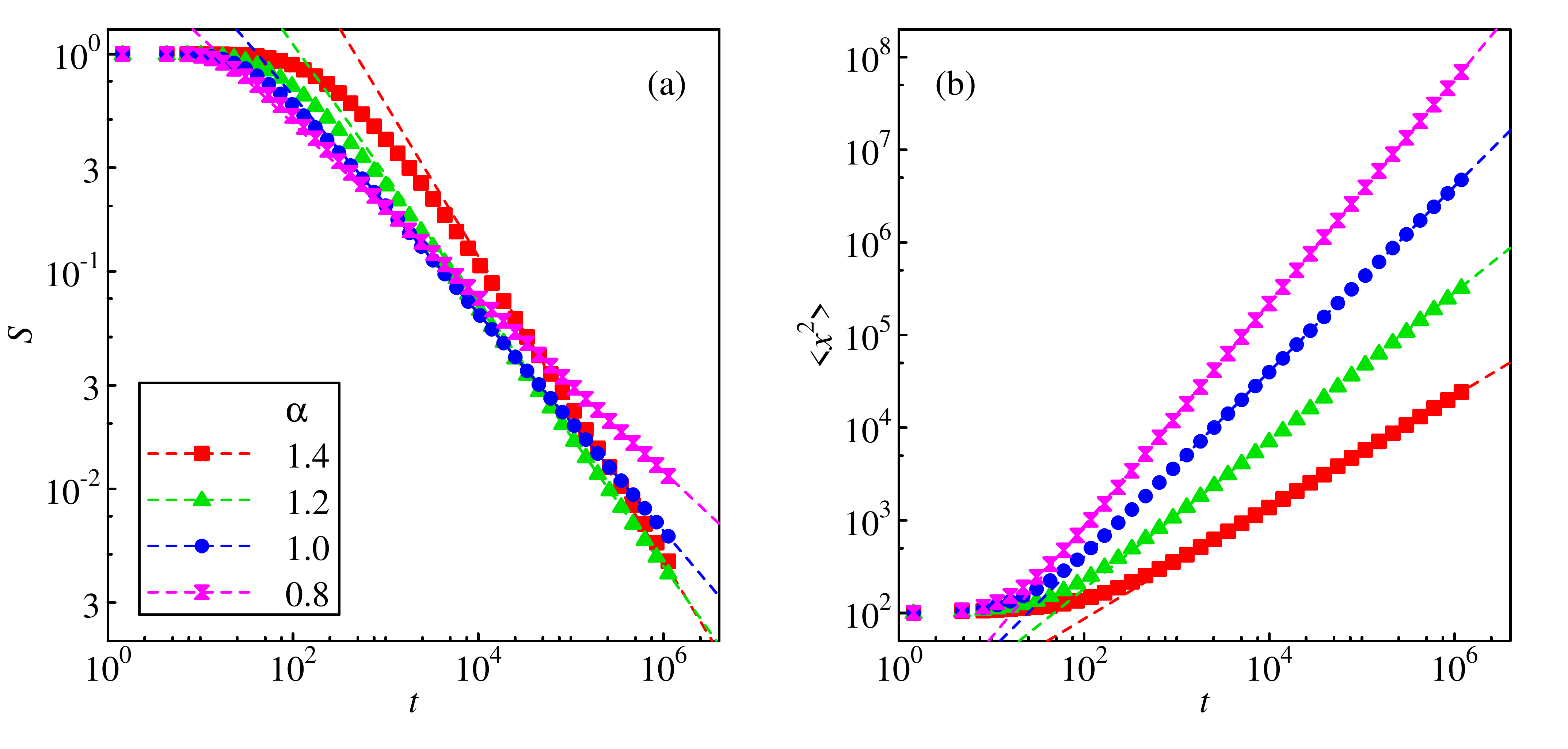}}
\caption{FLE on the semi-infinite interval $[0,\infty)$.
(a) Survival probability $S$ vs.\ time $t$ for several values of $\alpha$.
The initial particle number is $4\times 10^6$, the time step $\epsilon=0.08$, all particles start at $x_0=10$.
The dashed lines are fits of the long-time behavior to the
power law $S(t) \sim t^{-\alpha/2}$.
(b) Mean square displacement $\langle x^2 \rangle$ of the surviving particles vs.\ time $t$
for the same simulations as in panel (a).
The dashed lines are fits of the long-time behavior to the
power law $\langle x^2 \rangle \sim t^{2-\alpha}$.
The statistical errors of all data points are smaller than the symbol size.}
\label{fig:fle_nn_x2_all}
\end{figure}

We start by considering the FLE on a semi-infinite interval $[0,\infty)$ with
an absorbing wall at the origin. Particles start from rest at a position $x_0 > 0$
at time $t=0$.
Figure \ref{fig:fle_nn_x2_all}
presents the survival probability $S(t)$ and the mean-square displacement $\langle x^2 \rangle$
of the surviving particles for $x_0=10$ and several values of the exponent $\alpha$
that characterizes the covariance (\ref{eq:FGN_cov}) of the random forces in the FLE.
These correlations are persistent for $\alpha > 1$. Values $\alpha < 1$ correspond to antipersistent
noise which can be included in our discretized FLE based on the discussion in Sec.\ \ref{subsec:FLE_antipersistent}.

Figure \ref{fig:fle_nn_x2_all} resembles the corresponding Fig.\ \ref{fig:fbm_nn_x2_all}
for FBM. The order of the curves is
reversed, however, because persistent noise leads to subdiffusion for the FLE whereas antipersistent noise
produces superdiffusion (the opposite behavior compared to FBM).
Panel (a) demonstrates that the long-time decay of the survival probability follows the
power-law $S(t) \sim t^{-\theta}$ with $\theta=\alpha/2$. The mean-square displacement from the origin
for the surviving particles, shown in panel (b) follows the anomalous diffusion
relation $\langle x^2 \rangle \sim t^{2-\alpha}$.
Both power-law time dependencies agree with the corresponding FBM power laws if one
replaces the FBM anomalous diffusion exponent $\alpha$ with its FLE counterpart $2-\alpha$.

\begin{figure}
\centerline{\includegraphics[width=16cm]{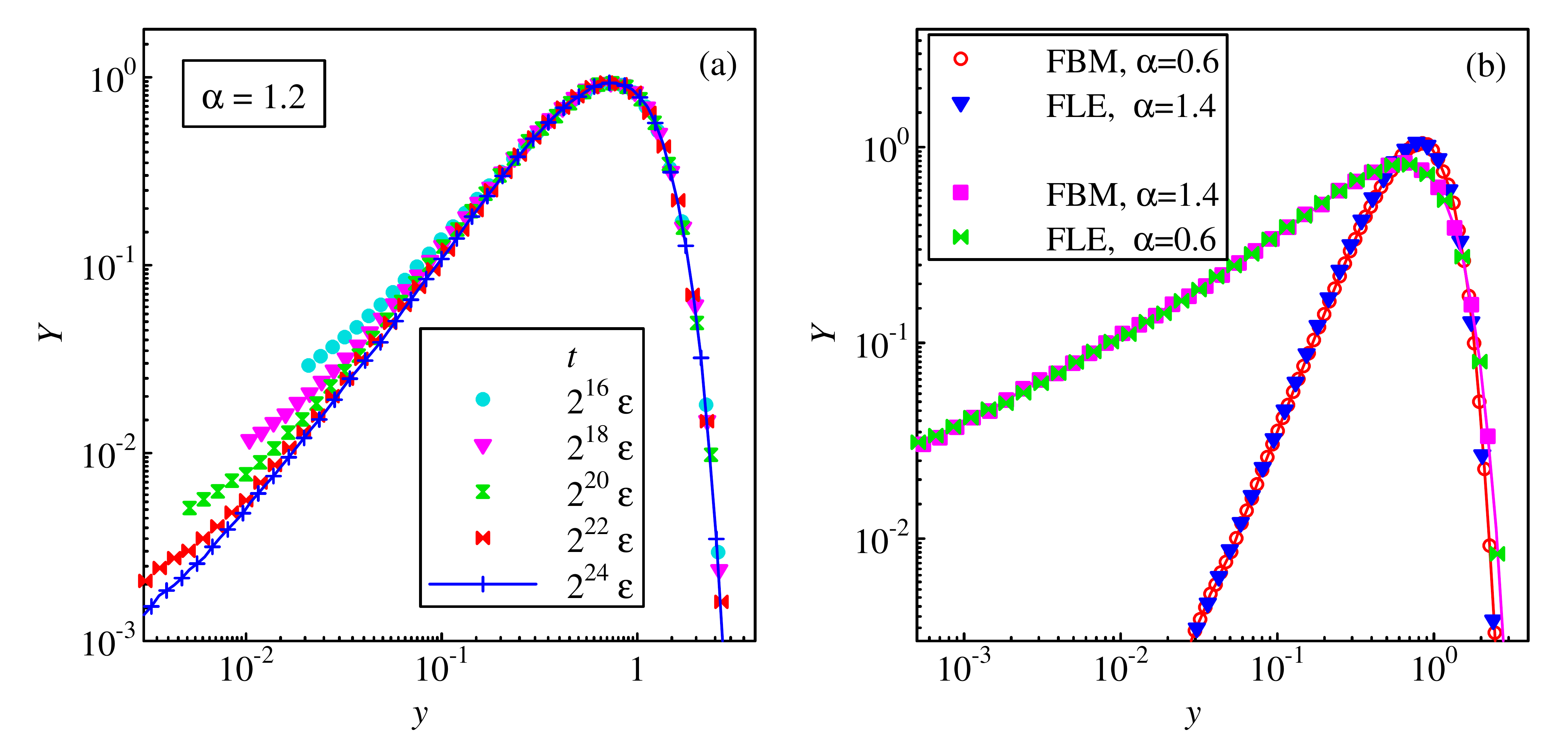}}
\caption{(a) Scaling plot of the conditional probability densities of the FLE on the interval $[0,\infty)$
at different times, expressed in terms of the variables $y=x/x_m(t)$ and $Y = P_s \, x_m(t)$ with $x_m(t)=\langle x^2(t)\rangle^{1/2}$.
The time step is $\epsilon=0.08$, and $\alpha=1.2$. The initial particle number is $4 \times 10^6$.
$P_s$ is averaged over a short time interval around the given time.
(b) Comparison of the scaling functions $Y^\mathrm{fbm}(y)$ for FBM and $Y^\mathrm{fle}(y)$ for
the FLE for two values of $\alpha$.
Data taken at time $2^{20}$ for FBM and $2^{24} \epsilon= 1\,342\,177$ for the FLE; the initial particle
number is at least $4\times 10^6$. The statistical errors of the data points are about a symbol size or
smaller.
}
\label{fig:FLE_distrib scaled}
\end{figure}

We now turn to the conditional probability density of the surviving particles, $P_s(x,t) = P(x,t)/S(t)$.
The simulations confirm that $P_s$ for the FLE fulfills the scaling form,
\begin{equation}
P_s(x,t) = \frac 1 {x_m(t)} Y_\alpha^\mathrm{fle} \left [ \frac x {x_m(t)}\right ]
\label{eq:Ps_scaling_fle}
\end{equation}
in the long-time limit  $x_m(t)  \gg x_0$. (Here,
$x_m(t)=\langle x^2(t)\rangle^{1/2}$, as before.)
Figure \ref{fig:FLE_distrib scaled}(a) demonstrates this for $\alpha=1.2$.
As was the case for FBM in Fig.\ \ref{fig:distrib_08_all_times_scaled}(a), the data collapse
is of good quality. Deviations of the small-$x$ data from the common master curve can be
attributed to finite-time discretization effects. They are suppressed with increasing $t$.

The scaling form (\ref{eq:Ps_scaling_fle}) for the FLE is identical to the corresponding
scaling form (\ref{eq:Ps_scaling_fbm}) for FBM.
It is interesting to compare the scaling functions $Y_\alpha^\mathrm{fbm}(y)$ for FBM and $Y_\alpha^\mathrm{fle}(y)$ for
the FLE. Figure \ref{fig:FLE_distrib scaled}(b) shows the scaled conditional probability densities
for both FBM and the FLE for two values of $\alpha$. (The data are taken at times $t>10^6$ so that
$x_m(t)  \gg x_0$, and time discretization artifacts can be neglected for the positions considered.)
Within our (small) statistical errors, the FBM and FLE scaling functions agree with each other if
one replaces $\alpha$ by $2-\alpha$,
\begin{equation}
Y_\alpha^\mathrm{fle}(y) = Y_{2-\alpha}^\mathrm{fbm}(y)~.
\label{eq:Yfle_Yfbm}
\end{equation}
This implies that the FLE scaling function is expected to follow the
power law $Y_\alpha^\mathrm{fle}(y) \sim y^\kappa$ with exponent $\kappa=\alpha/(2-\alpha)$
close to the absorbing wall, i.e. for $y \ll 1$.
Our numerical data can be fitted well with this power law all for all considered $\alpha$
(0.6, 0.8, 1.0, 1.2, and 1.4). For our most subdiffusive simulation, $\alpha=1.4$, some slight deviations
can be observed between this power law and the numerical data. As in the FBM case
in Sec.\ \ref{subsec:semi-infinite_FBM}, they can
be attributed to the simulations not having progressed far enough into the asymptotic
long-time regime.

%%%%%%%%%%%%%%%%%%%%%%%%%%%%%%%%%%%%%%%%%%%%%%%%%%%%%%%%%%%%%%%%%%%%%%%
\subsection{Stationary state on finite interval}
\label{subsec:finite_FLE}
%%%%%%%%%%%%%%%%%%%%%%%%%%%%%%%%%%%%%%%%%%%%%%%%%%%%%%%%%%%%%%%%%%%%%%%

To overcome the numerical difficulties associated with the particle loss occurring in the simulations on
the interval $[0,\infty)$ with an absorbing wall at the origin, we now consider the steady state of the FLE
on the finite interval $[-L/2,L/2]$ with absorbing walls at both ends and a particle source in the center.
This geometry is implemented analogously to the FBM case in Sec.\ \ref{subsec:finite_FBM}.
Particles start at the origin at $t=0$. When a particle is absorbed by one of the walls, it is
placed back at the origin. Unless noted otherwise, such a particle restarts
from the origin with a fresh set of random forces and without memory of its velocities before
the absorption event. For comparison, we also perform a number of calculations in which a particle
retains the velocity memory and noise sequence from before the absorption event.

\begin{figure}
\centerline{\includegraphics[width=16cm]{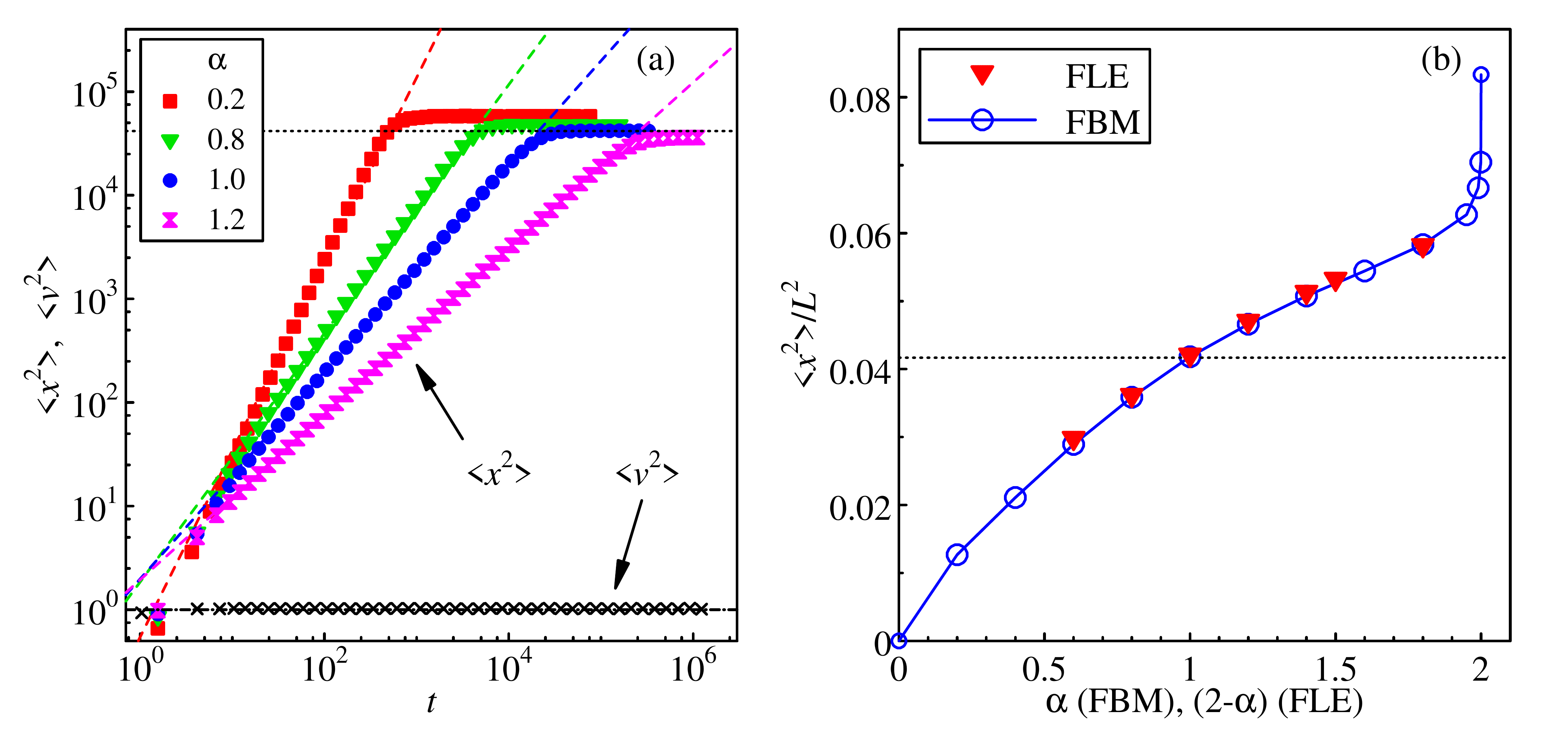}}
\caption{FLE on a finite interval $[-L/2,L/2]$ with absorbing walls at both ends and a source in the center.
(a) Mean square displacement $\langle x^2 \rangle$ vs.\ time $t$ for interval length $L=1000$ and several $\alpha$.
The time step is $\epsilon=0.04$, and the data are averages over $5\times 10^4$ to $10^5$ particles.
The dashed lines are fits to the anomalous diffusion power law $\langle x^2 \rangle \sim t^{2-\alpha}$.
The dotted line marks the stationary value $\langle x^2\rangle = L^2/24$ for normal diffusion
that follows from eq.\ (\ref{eq:P_stationary_normal}). Also shown is $\langle v^2 \rangle$ for $\alpha=1.2$.
The dash-dotted line shows the expected value of unity.
(b) Stationary value of $\langle x^2 \rangle/L^2$ vs.\ $\alpha$ or $2-\alpha$, comparing the FLE results to the findings
of Sec.\ \ref{subsec:finite_FBM} for FBM.  The FLE data are based on interval length $L=1000$ ($L=200$ for $\alpha=1.4$) and
$5\times 10^4$ to $10^5$ particles. $\langle x^2 \rangle$ is averaged over a number of time steps after
the stationary state is reached.
The dotted line marks the stationary value $\langle x^2\rangle = L^2/24$ for normal diffusion.
All statistical errors are much smaller than the symbol size.
}
\label{fig:fle_steady_overview}
\end{figure}

The time evolution of the mean-square displacement is shown in Fig.\ \ref{fig:fle_steady_overview}(a) for several values
of the exponent $\alpha$ that characterizes the random force covariance (\ref{eq:FGN_cov}).
The figure shows that the mean-square displacement grows ballistically for a brief
time initial period ($t \lessapprox 5$),
followed by anomalous diffusion governed by $\langle x^2 \rangle \sim t^{2-\alpha}$. After the particles
reach the wall, $\langle x^2 \rangle$ approaches a constant value that varies with $\alpha$,
just as in the FBM case. The mean-square velocity  $\langle v^2 \rangle$ quickly settles on a
value very close to unity, in agreement with the fluctuation dissipation theorem.
The slight deviation from unity (barely visible in the figure) can be attributed to the time
discretization error, as discussed in Sec.\ \ref{subsec:Simulation_FLE}.

Figure \ref{fig:fle_steady_overview}(b) shows how the stationary value of the scaled mean-square displacement
$\langle x^2 \rangle/L^2$ depends on the exponent $\alpha$, comparing our findings for the FLE with
the FBM results of Sec.\ \ref{subsec:finite_FBM}. The figure demonstrates that the stationary mean-square displacements
of FBM and the FLE agree nearly perfectly, if one replaces $\alpha$ by $2-\alpha$. This means that
$\langle x^2 \rangle = L^2/24$ for normal diffusion, $\alpha=1$ [see eq.\ (\ref{eq:P_stationary_normal})].
It also implies that $\langle x^2 \rangle$ approaches zero in most subdiffusive limit ($\alpha \to 2$ for the FLE), whereas it approaches the
value $L^2/12$ in the ballistic limit ($\alpha \to 0$ in the FLE).

\begin{figure}
\centerline{\includegraphics[width=16cm]{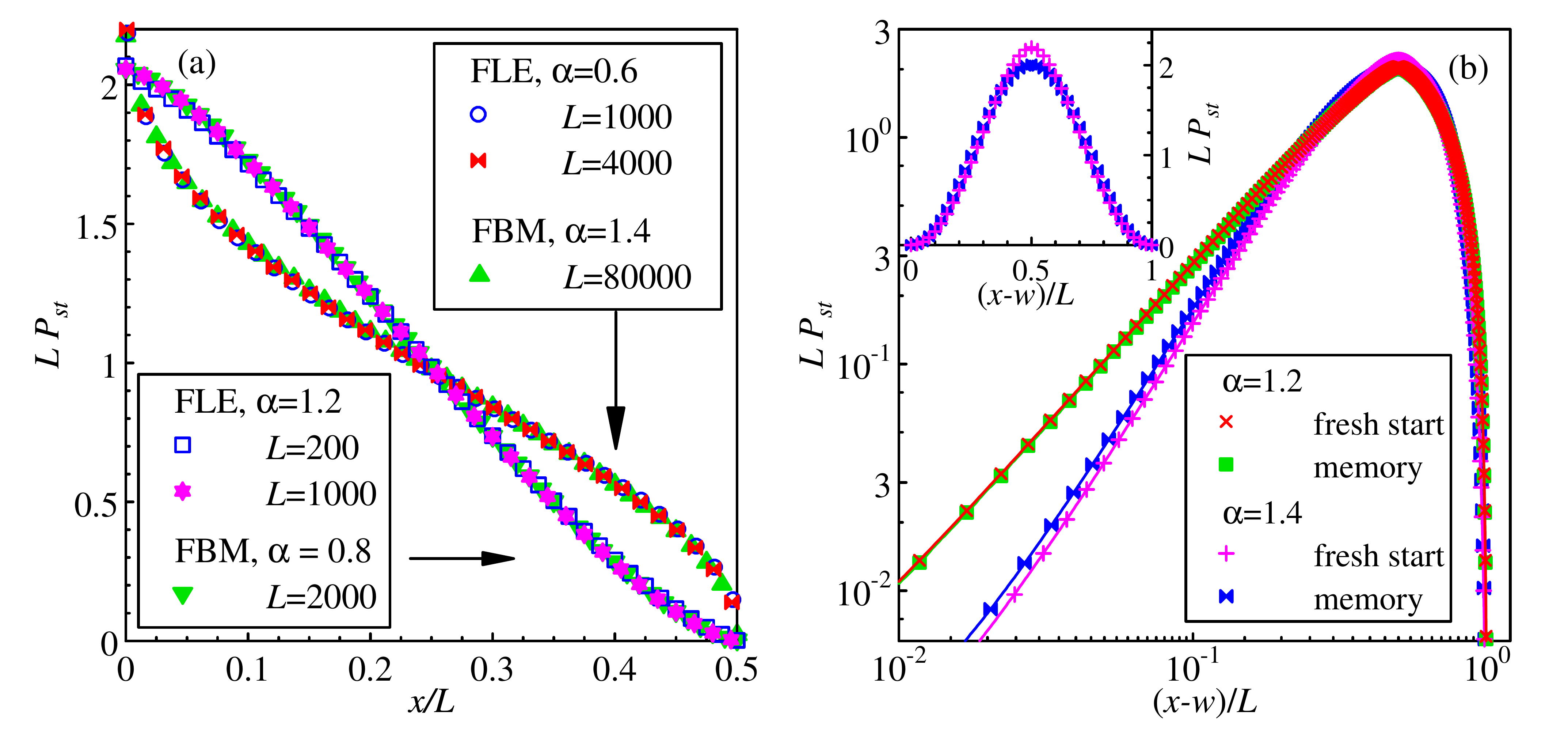}}
\caption{(a) Scaling plot of the stationary probability density for the FLE on the interval $[-L/2,L/2]$ showing $L P_{st}$ vs.\ $x/L$
for $\alpha=1.2$ and $\alpha=0.6$, $\epsilon=0.04$, and several interval length $L$ (only the right half of the interval is shown).
Each distribution is based
on $10^4$ to $10^5$ particles; $P_{st}$ is averaged over a number of time steps after a stationary state
has been reached. Also shown is the stationary probability density for FBM for $\alpha=0.8$ and 1.4.
(b) Effect of the restart condition on the stationary probability density for $\alpha=1.2$ ($L=1000$)
and $\alpha=1.4$ ($L=200$). The main panel shows $L P_{st}$ vs.\ the scaled distance $(x-w)/L$
from the absorbing wall at $-L/2$. The inset shows the data for $\alpha=1.4$ with linear axes.
All statistical errors are much smaller than the symbol size.
}
\label{fig:fle_distrib_scaled_interval}
\end{figure}

Let us now turn to the behavior of the stationary probability density $P_{st}(x,L)$ on the finite interval. By comparing the
probability densities for several interval length $L$, we confirm that $P_{st}(x,L)$ fulfills the scaling form
\begin{equation}
P_{st}(x,L) = \frac 1 L Z_\alpha^\mathrm{fle} \left ( \frac x L  \right)~,
\label{eq:P_scaling_interval_fle}
\end{equation}
where $Z_\alpha^\mathrm{fle}(z)$ is a dimensionless scaling function, in analogy to
eq.\ (\ref{eq:P_scaling_interval_fbm}) for the FBM case.
Figure \ref{fig:fle_distrib_scaled_interval}(a)  presents the corresponding scaling
plots for two values of $\alpha$.
The figure also shows that the scaled probability density of FBM with exponent $2-\alpha$
nearly perfectly agrees with the scaled probability density of the FLE with exponent
$\alpha$. This implies that the scaling functions fulfill the relation
\begin{equation}
Z_\alpha^\mathrm{fle}(z) = Z_{2-\alpha}^\mathrm{fbm}(z)~.
\label{eq:Zfle_Zfbm}
\end{equation}

In all FLE simulations on the finite interval reported so far, a particle absorbed by one of the walls
restarts from the center with a fresh set of random forces and no memory of its velocities before the absorption
event. To explore the effects of different restart conditions, we also perform a number of test calculations
in which particles retain their velocity history and the random force sequence after an absorption event.
A comparison of the stationary probability densities for the two restart conditions is
presented in Fig.\ \ref{fig:fle_distrib_scaled_interval}(b).
As anticipated in Sec.\ \ref{subsec:Simulation_FLE}, the stationary states for the two cases
are not exactly identical (see the inset of the figure). The difference is easy to understand at a qualitative
level. If a particle starts afresh with zero velocity, it will stay in the vicinity of the origin
for a longer time, increasing the probability density in this region. This effect appears to
increase with increasing $\alpha$, perhaps because the damping kernel ${\mathcal K}_{n-m}$
decays more slowly with time for larger $\alpha$. However, the main panel of
Fig.\ \ref{fig:fle_distrib_scaled_interval}(b)
demonstrates that the functional form of the probability density close to
the absorbing wall is not affected by the choice of restart condition. The curves for
``fresh restarts'' and ``retained history'' are parallel to each other for $x-w \ll L$,
i.e., they follow the same power law but with slightly different prefactors.

\begin{figure}
\centerline{\includegraphics[width=11cm]{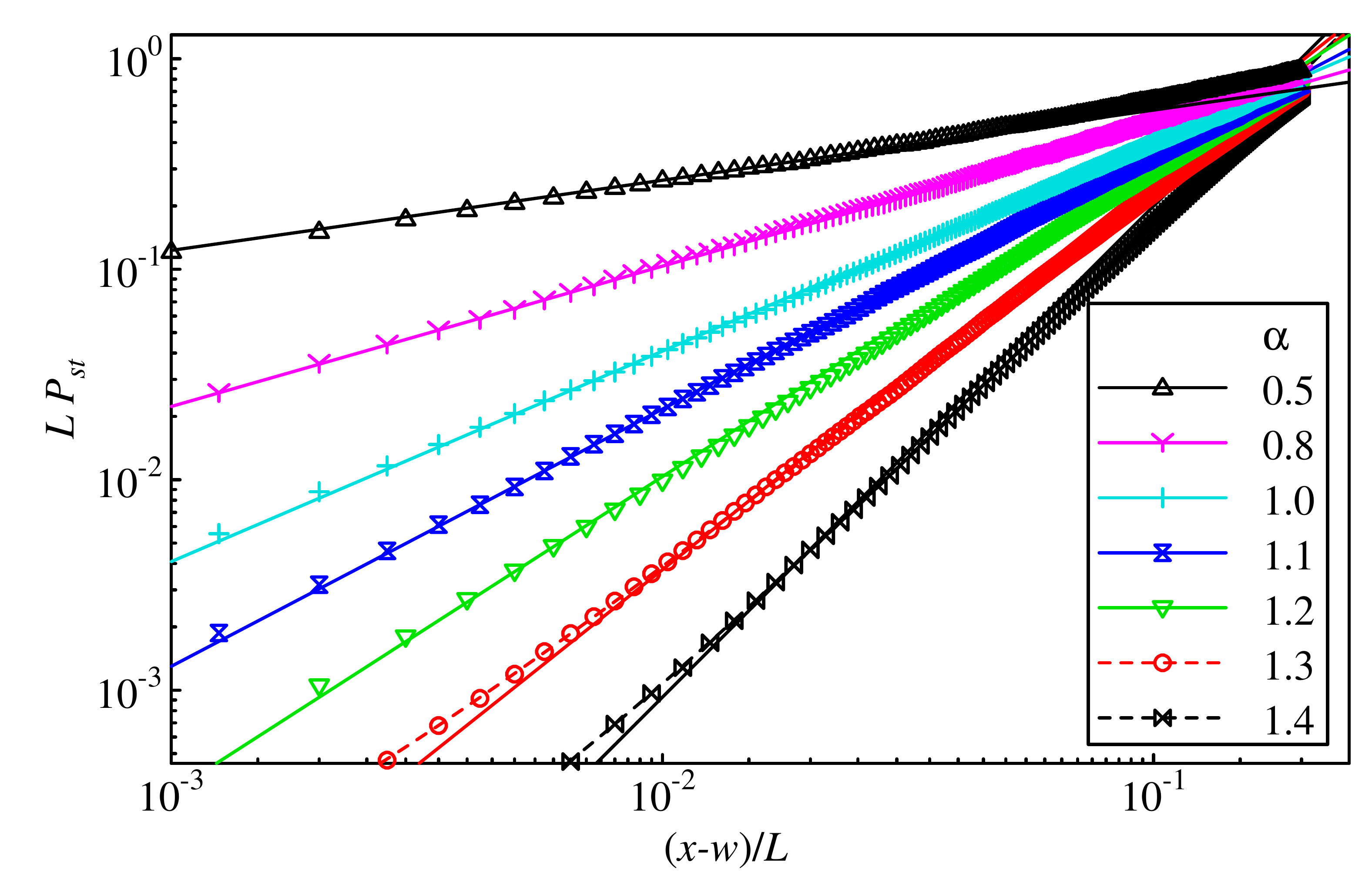}}
\caption{Scaled probability density $L P_{st}$ vs.\ scaled distance from the wall $(x-w)/L$ for the FLE on the interval $[-L/2,L/2]$
with $L$ ranging from 1000 for $\alpha=1.4$ to $10^5$ for $\alpha=0.5$. Each simulation employs 5000 to 20000
particles that perform
$2^{29}$ time steps of size $\epsilon=0.08$ ($2^{28}$ steps with $\epsilon=0.04$ for $\alpha=0.5$ and 0.8).
$P_{st}$ is averaged over a number of time steps after a stationary state
has been reached. The solid lines represent fits of the probability density in the wall region $(x-w)/L \lessapprox 0.03$
to the power law $P_{st} \sim (x-w)^\kappa$ with $\kappa= \alpha/(2-\alpha)$.
The dashed lines for $\alpha=0.6$ and 0.7 are fits with the function $P_{st} \sim (x-w+c)^\kappa$ which includes a subleading
finite-size correction $c$.
Statistical errors are much smaller than the symbol size.
}
\label{fig:fle_distrib_interval_zoom}
\end{figure}

To analyze the probability density close to the absorbing wall in more detail,
we have therefore performed a set of simulations using the "retained history"
restart condition. They are somewhat more efficient computationally because they do not
require the recalculation of a full random number sequence after each absorption event.
This allows us to reach up to $2^{29}$ time steps. Figure \ref{fig:fle_distrib_interval_zoom}
presents the resulting stationary probability density in the wall region for several
values of $\alpha$.
Sufficiently close to the absorbing wall, all distributions behave as powers of the distance $x-w$ (with $w=-L/2$)
from the wall. In this regime, the data
can be fitted with high accuracy with the expression $P_{st} \sim (x-w)^\kappa$ with $\kappa= \alpha/(2-\alpha)$.
(The exponent $\kappa$ agrees with the FBM conjecture \cite{ZoiaRossoMajumdar09}, $\kappa=2/\alpha-1$,
if $\alpha$ is replaced by $2-\alpha$, as was the case for the FLE on the semi-infinite interval.)
For our most subdiffusive simulations, $\alpha=1.4$ and 1.3, slight deviations between the power law and
the numerical data can be observed very close to the wall. They can be attributed to finite-size effects
as the interval length we reach in these subdiffusive simulations are still only moderately large.
In fact, the data can be fitted perfectly with the functional form $P_{st} \sim (x-w+c)^\kappa$
where the constant $c$ acts as a subleading finite-size correction.

%%%%%%%%%%%%%%%%%%%%%%%%%%%%%%%%%%%%%%%%%%%%%%%%%%%%%%%%%%%%%%%%%%%%%%%
\section{Conclusions}
\label{sec:conclusions}
%%%%%%%%%%%%%%%%%%%%%%%%%%%%%%%%%%%%%%%%%%%%%%%%%%%%%%%%%%%%%%%%%%%%%%%

In summary, we have employed large-scale computer simulations to study and compare the probability
density of FBM and the FLE in the presence of absorbing walls. We have considered two geometries,
(i) the spreading for particles from a fixed starting point $x_0>0$
on the semi-infinite interval $[0,\infty)$
with an absorbing wall at the origin and (ii) the steady state on a finite interval with
absorbing walls at both ends and a source in the center.

The FBM simulations on the semi-infinite interval agree with previous perturbative and 
numerical results in the literature \cite{ZoiaRossoMajumdar09,WieseMajumdarRosso11,DelormeWiese15,DelormeWiese16,ArutkinWalterWiese20,WalterWiese20}.
Specifically, the survival
probability decays as $S(t) \sim t^{\alpha/2-1}$ for long times, $2Kt^\alpha \gg x_0^2$.
In this limit, the conditional probability density for the surviving particles takes the scaling
form $P_s(x,t) =  Y_\alpha^\mathrm{fbm} [x/x_m(t)]\, /\,x_m(t)$ where $x_m(t) \sim t^{\alpha/2}$ is the root mean-square
displacement of the surviving particles. The dimensionless scaling function $Y_\alpha^\mathrm{fbm}(y)$
vanishes as $y^\kappa$ with $\kappa=2/\alpha-1$ close to the absorbing wall, i.e., for $y \ll 1$.

Interestingly, we have found the properties of the FLE on the semi-infinite interval to be identical to those of FBM
if one replaces the FBM anomalous diffusion exponent $\alpha$ by the anomalous diffusion exponent
$2-\alpha$ of the FLE. Specifically, the survival probability decays as $S(t) \sim t^{-\alpha/2}$
for long times whereas the root mean-square displacement of the surviving particles
increases as $x_m(t) \sim t^{1-\alpha/2}$. Moreover, the entire scaling functions of the conditional
probability densities for FBM and the FLE
map onto each other under this replacement, $Y_\alpha^\mathrm{fle}(y) = Y_{2-\alpha}^\mathrm{fbm}(y)$.
This implies that $Y_\alpha^\mathrm{fle}(y)$ vanishes for $y \to 0$
as $y^\kappa$ with $\kappa = \alpha/(2-\alpha)$.

Our results for the steady state on the finite interval $[-L/2,L/2]$ paint a similar picture.
The stationary probability density depends on $\alpha$ and fulfills the scaling form
$P_{st}(x,L) = Z_\alpha^\mathrm{fbm}(x/L)\, / \,L$ for FBM, and analogously for the FLE.
Close to the absorbing walls at positions $w=\pm L/2$, the FBM scaling function behaves
as $Z_\alpha^\mathrm{fbm}(z) \sim |z\mp 1/2|^\kappa$,
with the same exponent $\kappa=2/\alpha-1$ as for the semi-infinite
interval. The behavior of the FLE in this geometry is identical to that of FBM
after one replaces $\alpha$ by $2-\alpha$. The scaling functions thus fulfill
the relation $Z_\alpha^\mathrm{fle}(z) = Z_{2-\alpha}^\mathrm{fbm}(z)$, and
$Z_\alpha^\mathrm{fle}(z)$ vanishes with exponent $\kappa= \alpha/(2-\alpha)$ close
to $z=\pm 1/2$.

It is interesting to compare the effects of absorbing and reflecting boundaries on the
probability density. For FBM, there is a clear analogy between the two cases.
The probability density vanishes as $P \sim x^{2/\alpha-1}$ with distance $x$ from
an absorbing wall whereas it behaves as $P \sim x^{2/\alpha-2}$ close to a reflecting wall
\cite{WadaVojta18,WadaWarhoverVojta19,Guggenbergeretal19,VHSJGM20}.
This means, for both boundary conditions, persistent noise increases the probability
density close to the wall compared to normal diffusion. Anti-persistent noise, in contrast,
reduces the probability density compared to normal diffusion.

For the FLE, such a clear analogy between absorbing and reflecting boundary conditions does
not exist. The probability density of the FLE with absorbing walls can be mapped onto that
of FBM with absorbing walls by replacing $\alpha$ with $2-\alpha$, implying that the functional
form of the probability density is strongly $\alpha$ dependent. In contrast, the
stationary distribution of the FLE on a finite interval with reflecting walls is
completely uniform independent of the value of $\alpha$ (as required in thermal equilibrium),
and any accumulation and depletion effects for the FLE on a semi-infinite interval
with a reflecting wall at the origin are much weaker than those of FBM \cite{VojtaSkinnerMetzler19}.

What is the reason that FBM and the FLE behave very similarly for absorbing boundaries
but in qualitatively different ways for reflecting boundaries? In the absorbing case,
the particles do not actually ``interact'' with the boundary; whenever a particle
crosses the boundary it is removed. This means that the free FBM or FLE trajectories
completely determine the fate of a particle in the absorbing case.
In the reflecting case, a particle stays in the system after reaching the wall,
and its subsequent behavior depends on the wall interaction. For FBM, the
(putative) displacements are determined by the noise and thus not affected by the wall.
For the FLE, in contrast, the noise defines the forces while the displacements
are determined by the instantaneous velocity which is affected by the wall.

We conclude by emphasizing that in many applications, perfectly absorbing or reflecting boundaries
should be considered limiting cases, and the actual interaction of the trajectories
with the confining walls may be more complicated. Moreover, the long-range power-law correlations
are often regularized beyond some time or length scale.  To account for such a cutoff of the correlations,
one can employ tempered fractional Gaussian noise \cite{MolinaGarciaetal18}.
Exploring these more complex situations remains a task for the future.

%%%%%%%%%%%%%%%%%%%%%%%%%%%%%%%%%%%%%%%%%%%%%%%%%%%%%%%%%%%%%%%%%%%%%%%
\section{Acknowledgments}
%%%%%%%%%%%%%%%%%%%%%%%%%%%%%%%%%%%%%%%%%%%%%%%%%%%%%%%%%%%%%%%%%%%%%%%

This work was supported in part by a Cottrell SEED award from Research Corporation and by the
National Science Foundation under Grants No.\ DMR-1828489 and No.\ OAC-1919789. We thank
R. Metzler for valuable discussions.

%\addcontentsline{toc}{section}{References}
%\section{References}

\small

\bibliography{../00Bibtex/rareregions}

\providecommand{\newblock}{}
\begin{thebibliography}{10}
\expandafter\ifx\csname url\endcsname\relax
  \def\url#1{{\tt #1}}\fi
\expandafter\ifx\csname urlprefix\endcsname\relax\def\urlprefix{URL }\fi
\providecommand{\eprint}[2][]{\url{#2}}
% Bibliography created with iopart-num v2.1
% /biblio/bibtex/contrib/iopart-num

\bibitem{Einstein_book56}
Einstein A 1956 {\em Investigations on the Theory of the Brownian Movement\/}
  (New York: Dover)

\bibitem{Smoluchowski18}
von Smoluchowski M 1918 Versuch einer mathematischen {T}heorie der
  {K}oagulationskinetik kolloider {L\"o}sungen {\em Z. Phys. Chem.\/} {\bf 92U}
  129

\bibitem{Langevin08}
Langevin P 1908 Sur la th\'eorie du mouvement brownien {\em C. R. Acad. Sci.
  Paris\/} {\bf 146} 530--533

\bibitem{Hughes95}
Hughes B 1995 {\em Random Walks and Random Environments, Volume 1: Random
  Walks\/} (Oxford: Oxford University Press)

\bibitem{MetzlerKlafter00}
Metzler R and Klafter J 2000 The random walk's guide to anomalous diffusion: a
  fractional dynamics approach {\em Physics Reports\/} {\bf 339} 1 -- 77

\bibitem{HoeflingFranosch13}
H{\"o}fling F and Franosch T 2013 Anomalous transport in the crowded world of
  biological cells {\em Rep. Progr. Phys.\/} {\bf 76} 046602

\bibitem{BressloffNewby13}
Bressloff P~C and Newby J~M 2013 Stochastic models of intracellular transport
  {\em Rev. Mod. Phys.\/} {\bf 85}(1) 135--196

\bibitem{MJCB14}
Metzler R, Jeon J~H, Cherstvy A~G and Barkai E 2014 Anomalous diffusion models
  and their properties: non-stationarity{,} non-ergodicity{,} and ageing at the
  centenary of single particle tracking {\em Phys. Chem. Chem. Phys.\/} {\bf
  16}(44) 24128--24164

\bibitem{MerozSokolov15}
Meroz Y and Sokolov I~M 2015 A toolbox for determining subdiffusive mechanisms
  {\em Physics Reports\/} {\bf 573} 1 -- 29 ISSN 0370-1573

\bibitem{MetzlerJeonCherstvy16}
Metzler R, Jeon J~H and Cherstvy A 2016 Non-brownian diffusion in lipid
  membranes: Experiments and simulations {\em Biochimica et Biophysica Acta\/}
  {\bf 1858} 2451 -- 2467 ISSN 0005-2736

\bibitem{XCLLL08}
Xie X~S, Choi P~J, Li G~W, Lee N~K and Lia G 2008 Single-molecule approach to
  molecular biology in living bacterial cells {\em Annual Review of
  Biophysics\/} {\bf 37} 417--444

\bibitem{BrauchleLambMichaelis12}
Br{\"a}uchle C, Lamb D~C and Michaelis J 2012 {\em Single Particle Tracking and
  Single Molecule Energy Transfer\/} (Weinheim: Wiley-VCH)

\bibitem{ManzoGarciaParajo15}
Manzo C and Garcia-Parajo M~F 2015 A review of progress in single particle
  tracking: from methods to biophysical insights {\em Rep. Progr. Phys.\/} {\bf
  78} 124601

\bibitem{SzymanskiWeiss09}
Szymanski J and Weiss M 2009 Elucidating the origin of anomalous diffusion in
  crowded fluids {\em Phys. Rev. Lett.\/} {\bf 103}(3) 038102

\bibitem{MWBK09}
Magdziarz M, Weron A, Burnecki K and Klafter J 2009 Fractional brownian motion
  versus the continuous-time random walk: A simple test for subdiffusive
  dynamics {\em Phys. Rev. Lett.\/} {\bf 103}(18) 180602

\bibitem{WeberSpakowitzTheriot10}
Weber S~C, Spakowitz A~J and Theriot J~A 2010 Bacterial chromosomal loci move
  subdiffusively through a viscoelastic cytoplasm {\em Phys. Rev. Lett.\/} {\bf
  104}(23) 238102

\bibitem{Jeonetal11}
Jeon J~H, Tejedor V, Burov S, Barkai E, Selhuber-Unkel C, Berg-S\o{}rensen K,
  Oddershede L and Metzler R 2011 In vivo anomalous diffusion and weak
  ergodicity breaking of lipid granules {\em Phys. Rev. Lett.\/} {\bf 106}(4)
  048103

\bibitem{JMJM12}
Jeon J~H, Monne H~M~S, Javanainen M and Metzler R 2012 Anomalous diffusion of
  phospholipids and cholesterols in a lipid bilayer and its origins {\em Phys.
  Rev. Lett.\/} {\bf 109}(18) 188103

\bibitem{Tabeietal13}
Tabei S~M~A, Burov S, Kim H~Y, Kuznetsov A, Huynh T, Jureller J, Philipson L~H,
  Dinner A~R and Scherer N~F 2013 Intracellular transport of insulin granules
  is a subordinated random walk {\em Proc. Nat. Acad. Sci.\/} {\bf 110}
  4911--4916 ISSN 0027-8424

\bibitem{ChakravartiSebastian97}
Chakravarti N and Sebastian K 1997 Fractional brownian motion models for
  polymers {\em Chem. Phys. Lett.\/} {\bf 267} 9 -- 13 ISSN 0009-2614

\bibitem{Panja10}
Panja D 2010 Generalized langevin equation formulation for anomalous polymer
  dynamics {\em J. Stat. Mech.\/} {\bf 2010} L02001

\bibitem{MRRS02}
Mikosch T, Resnick S, Rootzen H and Stegeman A 2002 Is network traffic
  appriximated by stable levy motion or fractional brownian motion? {\em Ann.
  Appl. Probab.\/} {\bf 12} 23--68

\bibitem{JanusonisDeteringMetzlerVojta20}
Janu{\v s}onis S, Detering N, Metzler R and Vojta T 2020 Serotonergic axons as
  fractional brownian motion paths: Insights into the self-organization of
  regional densities {\em Front. Comp. Neuroscience\/} {\bf 14} 56 ISSN
  1662-5188

\bibitem{ComteRenault98}
Comte F and Renault E 1998 Long memory in continuous-time stochastic volatility
  models {\em Math. Financ.\/} {\bf 8} 291

\bibitem{RostekSchoebel13}
Rostek S and Sch{\"o}bel R 2013 A note on the use of fractional brownian motion
  for financial modeling {\em Econom. Model.\/} {\bf 30} 30 -- 35 ISSN
  0264-9993

\bibitem{Kolmogorov40}
Kolmogorov A~N 1940 Wienersche spiralen und einige andere interessante kurven
  im hilbertschen raum {\em C. R. (Doklady) Acad. Sci. URSS (N.S.)\/} {\bf 26}
  115--118

\bibitem{MandelbrotVanNess68}
Mandelbrot B~B and Ness J~W~V 1968 Fractional brownian motions, fractional
  noises and applications {\em SIAM Review\/} {\bf 10} 422--437

\bibitem{Kahane85}
Kahane J~P 1985 {\em Some Random Series of Functions\/} (London: Cambridge
  University Press)

\bibitem{Yaglom87}
Yaglom A~M 1987 {\em Correlation Theory of Stationary and Related Random
  Functions\/} (Heidelberg: Springer)

\bibitem{Beran94}
Beran J 1994 {\em Statistics for Long-Memory Processes\/} (New York: Chapman \&
  Hall)

\bibitem{BHOZ08}
Biagini F, Hu Y, {\O}ksendal B and Zhang T 2008 {\em Stochastic Calculus for
  Fractional Brownian Motion and Applications\/} (Berlin: Springer)

\bibitem{Redner_book01}
Redner S 2001 {\em A guide to first-passage processes\/} (Cambridge: Cambridge
  University Press)

\bibitem{HansenEngoyMaloy94}
Hansen A, Eng{\o}y T and M{\aa}l{\o}y K~J 1994 Measuring hurst exponents with
  the first return method {\em Fractals\/} {\bf 02} 527--533

\bibitem{DingYang95}
Ding M and Yang W 1995 Distribution of the first return time in fractional
  brownian motion and its application to the study of on-off intermittency {\em
  Phys. Rev. E\/} {\bf 52}(1) 207--213

\bibitem{KKMCBS97}
Krug J, Kallabis H, Majumdar S~N, Cornell S~J, Bray A~J and Sire C 1997
  Persistence exponents for fluctuating interfaces {\em Phys. Rev. E\/} {\bf
  56}(3) 2702--2712

\bibitem{Molchan99}
Molchan G~M 1999 Maximum of a fractional brownian motion: Probabilities of
  small values {\em Commun. Math. Phys.\/} {\bf 205} 97--111

\bibitem{JeonChechkinMetzler11}
Jeon J~H, Chechkin A~V and Metzler R 2011 First passage behaviour of fractional
  brownian motion in two-dimensional wedge domains {\em {EPL} (Europhysics
  Letters)\/} {\bf 94} 20008

\bibitem{AurzadaLifshits19}
Aurzada F and Lifshits M~A 2019 The first exit time of fractional brownian
  motion from a parabolic domain {\em Theory Probab. Appl.\/} {\bf 64} 490--497

\bibitem{Klimontovich_book95}
Klimontovich Y~L 1995 {\em Statistical theory of open systems - Volume 1: A
  unified approach to kinetic description of processes in active systems\/}
  (Dordrecht: Kluwer Academic Publishers)

\bibitem{Kubo66}
Kubo R 1966 The fluctuation-dissipation theorem {\em Rep. Progr. Phys.\/} {\bf
  29} 255--284

\bibitem{Lutz01}
Lutz E 2001 Fractional langevin equation {\em Phys. Rev. E\/} {\bf 64}(5)
  051106

\bibitem{WadaVojta18}
Wada A~H~O and Vojta T 2018 Fractional brownian motion with a reflecting wall
  {\em Phys. Rev. E\/} {\bf 97}(2) 020102

\bibitem{WadaWarhoverVojta19}
Wada A~H~O, Warhover A and Vojta T 2019 Non-gaussian behavior of reflected
  fractional brownian motion {\em J. Stat. Mech.\/} {\bf 2019} 033209

\bibitem{Guggenbergeretal19}
Guggenberger T, Pagnini G, Vojta T and Metzler R 2019 Fractional brownian
  motion in a finite interval: correlations effect depletion or accretion zones
  of particles near boundaries {\em New J. Phys.\/} {\bf 21} 022002

\bibitem{VHSJGM20}
Vojta T, Halladay S, Skinner S, Janu\ifmmode~\check{s}\else \v{s}\fi{}onis S,
  Guggenberger T and Metzler R 2020 Reflected fractional brownian motion in one
  and higher dimensions {\em Phys. Rev. E\/} {\bf 102}(3) 032108

\bibitem{VojtaSkinnerMetzler19}
Vojta T, Skinner S and Metzler R 2019 Probability density of the fractional
  langevin equation with reflecting walls {\em Phys. Rev. E\/} {\bf 100}(4)
  042142

\bibitem{ChatelainKantorKardar08}
Chatelain C, Kantor Y and Kardar M 2008 Probability distributions for polymer
  translocation {\em Phys. Rev. E\/} {\bf 78}(2) 021129

\bibitem{ZoiaRossoMajumdar09}
Zoia A, Rosso A and Majumdar S~N 2009 Asymptotic behavior of self-affine
  processes in semi-infinite domains {\em Phys. Rev. Lett.\/} {\bf 102}(12)
  120602

\bibitem{WieseMajumdarRosso11}
Wiese K~J, Majumdar S~N and Rosso A 2011 Perturbation theory for fractional
  brownian motion in presence of absorbing boundaries {\em Phys. Rev. E\/} {\bf
  83}(6) 061141

\bibitem{DelormeWiese15}
Delorme M and Wiese K~J 2015 Maximum of a fractional brownian motion: Analytic
  results from perturbation theory {\em Phys. Rev. Lett.\/} {\bf 115}(21)
  210601

\bibitem{DelormeWiese16}
Delorme M and Wiese K~J 2016 Perturbative expansion for the maximum of
  fractional brownian motion {\em Phys. Rev. E\/} {\bf 94}(1) 012134

\bibitem{ArutkinWalterWiese20}
Arutkin M, Walter B and Wiese K~J 2020 Extreme events for fractional brownian
  motion with drift: Theory and numerical validation {\em Phys. Rev. E\/} {\bf
  102}(2) 022102

\bibitem{Qian03}
Qian H 2003 Fractional brownian motion and fractional gaussian noise {\em
  Processes with Long-Range Correlations: Theory and Applications\/} ed
  Rangarajan G and Ding M (Berlin, Heidelberg: Springer) pp 22--33 ISBN
  978-3-540-44832-7

\bibitem{MHSS96}
Makse H~A, Havlin S, Schwartz M and Stanley H~E 1996 Method for generating
  long-range correlations for large systems {\em Phys. Rev. E\/} {\bf 53}(5)
  5445--5449

\bibitem{Lecuyer99}
L'Ecuyer P 1999 Tables of maximally equidistributed combined lfsr generators
  {\em Math. Comput.\/} {\bf 68} 261--269 ISSN 0025-5718

\bibitem{Marsaglia05}
Marsaglia G 2005 Double precision {RNG}s Posted to sci.math.num-analysis
  http://sci.tech-archive.net/Archive/sci.math.num-analysis/2005-11/msg00352.html
  \urlprefix\url{http://sci.tech-archive.net/Archive/sci.math.num-analysis/2005-11/msg00352.html}

\bibitem{Zwanzig_book01}
Zwanzig R 2001 {\em Nonequilibrium Statistical Mechanics\/} (Oxford: Oxford
  University Press)

\bibitem{Haenggi78}
H{\"a}nggi P 1978 Correlation functions and masterequations of generalized
  (non-markovian) langevin equations {\em Z. Phys. B\/} {\bf 31} 407--416 ISSN
  1431-584X

\bibitem{Goychuk12}
Goychuk I 2012 {\em Viscoelastic Subdiffusion: Generalized {L}angevin Equation
  Approach\/} (John Wiley \& Sons, Ltd) pp 187--253 ISBN 9781118197714

\bibitem{Wiese19}
Wiese K~J 2019 First passage in an interval for fractional brownian motion {\em
  Phys. Rev. E\/} {\bf 99}(3) 032106

\bibitem{WalterWiese20}
Walter B and Wiese K~J 2020 Sampling first-passage times of fractional brownian
  motion using adaptive bisections {\em Phys. Rev. E\/} {\bf 101}(4) 043312

\bibitem{MolinaGarciaetal18}
Molina-Garcia D, Sandev T, Safdari H, Pagnini G, Chechkin A and Metzler R 2018
  Crossover from anomalous to normal diffusion: truncated power-law noise
  correlations and applications to dynamics in lipid bilayers {\em New J.
  Phys.\/} {\bf 20} 103027

\end{thebibliography}
\bibliographystyle{iopart-num-title}
\end{document}